\documentclass{article}
\usepackage[dvips] {graphics}
\usepackage {epsfig}
\usepackage[mathscr]{eucal}
\usepackage{amsfonts}
\usepackage{amsmath}
\usepackage{pgf,xcolor}
\usepackage{dsfont}

\usepackage[american]{babel}
\usepackage{bm}

\newcommand{\bx}{\mathbf{x}}
\newcommand{\bk}{\mathbf{k}}
\newcommand{\eps}{\varepsilon}

\newcommand{\sv}{\, ,}

\newcommand{\BE}{\begin{equation}}
\newcommand{\EE}{\end{equation}}

\usepackage{geometry}
\begin{document}

\title{Planar phonon anisotropy, and a way to detect local equilibrium temperature in graphene}
\date{}

\author{Marco Coco\thanks{{\tt m.coco@univpm.it}\\
		Department of Industrial Engineering and Mathematical Sciences \\
		Polytechnic University of Marche\\
		via Brecce Bianche, 12, 60131, Ancona, Italy}}

\maketitle
\begin{abstract}
The effect of inclusion of the planar phonon anisotropy on thermo-electrical behavior of graphene is analyzed. Charge transport is simulated by means of Direct Simulation Monte Carlo technique coupled with numerical solution of the phonon Boltzmann equations based on deterministic methods. 

The definition of the crystal lattice local equilibrium temperature is investigated as well and the results furnish possible alternative approaches to identify it starting from measurements of electric current density, with relevant experimental advantages, {{which could help to overcome}} the present difficulties regarding thermal investigation of graphene. 

{{Positive implications are expected for many applications, as the field of electronic devices, which needs a coherent tool for simulation of charge and hot phonon transport; the correct definition of the local equilibrium temperature is in turn fundamental for the study, design and prototyping of cooling mechanisms for graphene-based devices.}}
\end{abstract}
\section{Introduction}
\label{Introduction}

Graphene has a wide range of interesting properties and it is the main
candidate for future nano-electronics and many other applications in several
fields. A lot of studies were devoted to the understanding of its electronic
properties and to the simulation methods for semiconductor applications.
Particular attention was paid by Monte Carlo techniques, which were already
established for semiconductor devices~\cite{LP,BJ,JacReg}, and above all
by the correct inclusion of the Pauli exclusion principle in the standard Ensemble Monte
Carlo approach~\cite{LuFe}, which in graphene is no longer negligible due to its
high electronic degeneracy; a new Direct Simulation Monte Carlo scheme (DSMC)
was developed in~\cite{RoMaCo} while  in the
literature~\cite{BoTho,BoAda,ZeBuEsSh,1} it remained open a discussion about the
classical liouvillian or quantum-like interpretation of the free flight
step~\cite{Tady}, whose study was addressed in~\cite{PRB}. 

Some other modeling aspects still deserve a better analysis in order to have a coherent and self-consistent tool for DSMC simulations of charge and phonon transport in graphene, in particular the role of planar phonon anisotropy and the definition of the crystal lattice local equilibrium temperature. The main aim of this paper is to contribute to fill this gap.

In the study of transport processes in semiconductor materials, phonons are
often distinguished just based on their branch, i.e.~acoustical and optical,
neglecting their polarization, longitudinal, transversal and flexural, {{which
		is taken into account only by introducing a degeneracy factor. Such an approach
		could lead to coarse approximations; for example, for a correct simulation of
		transport behavior of graphene when thermal effects are considered, the
		out-of-plane $Z$-phonons have to be included as a distinct population
		because, although they do not directly interact with electrons}}, their thermal
inertia, particularly for the flexural acoustic $ZA$ phonons, is not
negligible~\cite{CR_HT, MoMa, NiBa, Pop, Ruan, Bao}. However, the isotropic approximation, for
which the in-plane acoustic and optical phonons are considered as a unique
population, is often accepted for simulations {{(see for
		example~\cite{RoMaCo,IsKa,ThBa,CoMasRo,Yinga,Rengel1,Compel,Quispe},
		and references therein), and an exhaustive analysis of the implications of such
		a hypothesis is not present, at best of our knowledge, in the literature.}}

{{We will analyze the consequence of this assumption when phonons are taken at thermal bath and in the case in which the crystal heating is included, and we will also distinguish the effect of a complete phonon anisotropy and that due to only acoustical or optical phonon anisotropy, respectively}}. 

Fundamental aspect still matter of debate is the definition of the crystal
lattice local equilibrium temperature of graphene {{for which two main
		approaches are present in the literature; one is based on the properties of the
		phonon--phonon collision operator, the other on local thermodynamic equilibrium~\cite{Vas, London, Entropy, Hao, Lu, Peraud}. Interesting results will be derived
		investigating the effect of planar phonon anisotropy by using these two
		definitions of local equilibrium temperature}}. {{The lack of definitive
		experimental data has so far left the discussion about the definition of the local equilibrium temperature and other important thermal properties of graphene open; our work attempts to
		advance knowledge regarding this important issue, for which,  also from a
		theoretical point of view}}, there is still much to be studied, as for example
the definition and characterization of phonon relaxation times.

The phonon--phonon collisions in the phonon Boltzmann equations are a very
difficult part to treat both theoretically and numerically, and the
Bhatnagar--Gross--Krook (BGK) approximation is usually used; it describes the
tendency of each phonon branch to reach an equilibrium condition with a common
temperature which is the crystal lattice local equilibrium $T_{LE}$, and it
depends also on the relaxation times. Constant relaxation times are sometimes
considered, above all when the isotropic approximation for phonons is
embrached~\cite{CoMasRo,Rengel2}; more refined model is given by the
frequency-integrated temperature dependent relaxation times
in~\cite{Bao,Hao,Peraud}; new studies were made also on the inclusion
of the four--phonon scattering for the definition of relaxation times but
numerical data are available only at room temperature~\cite{Ruan,Feng1}. To
overcome such a difficulty {{we will analyze the sensitivity of our
		results to relaxation times}}; their influence and the convergence properties
of Monte Carlo procedure when hot phonons are considered were investigated
in~\cite{CocoJCTT}.

There are many studies in the literature about thermal conductivity and thermal behavior of graphene, but they are limited by practical difficulties of measurements. The results on the electric characteristic curves we obtain could lead to further developments about identification of the graphene local equilibrium temperature {{for which, as said, a unique definition is yet not clear, providing also an experimental way of investigation,}} based on measurements of the electric current density and not directly of the thermodynamic quantities, with relevant experimental advantages. 

{{The computational modeling of phonon's anisotropy addressed in this paper
		together with the correct inclusion of the Pauli principle~\cite{PRB} and the
		coupling of the DSMC simulations with deterministic solutions of the phonon
		Boltzmann equations~\cite{CR_HT,Coco_IJMPB} can constitute a complete and
		coherent simulation tool for the study of electrons and phonons transport
		processes in graphene. Its advantages could be extended to all applications
		dealing with electronic devices and also to other physical phenomena when such
		a kind of simulation's procedures are needed. }}

{{We will consider a single suspended layer of graphene, under the action of an external electric field, with infinite length in the direction transversal to it, in such a way to study homogeneous solutions and to not consider any effect due to the boundary conditions. Moreover, any dissipation mechanism is introduced and thermal equilibrium cannot be reached.}}

The plan of the paper is as follows. In Section \ref{Model} the mathematical model and the simulation approach are presented; Section \ref{results} is devoted to the discussion of numerical results and in Section \ref{Conclusions} we give conclusions.

\section{Mathematical model and simulation procedure}
\label{Model}

	Graphene is made of carbon atoms arranged in a honeycomb hexagonal
structure~\cite{CaNe}; the Brillouin zone $\mathcal B$ is hexagonal as well;
electrons are mainly located around its vertices called {\emph{Dirac points}},
pairwise equivalent, so it is enough to consider only two points $K$ and
$K'$; we study the electron dynamics in only one point $K$ and we
introduce a valley degeneracy $\alpha_v=2$ when necessary. In the numerical
calculations, we choose a Fermi level $\varepsilon_F$ high enough so that the
valence band is completely occupied by electrons and we can take into account
only the conduction band, as an n-type doping in traditional semiconductor
materials.

Graphene has a zero gap between the valence and the conduction bands and around Dirac points the
following dispersion relation holds~\cite{CaNe}:

\begin{equation}
	\eps({\bf k}) =   \hbar \, v_F \left| \bk -\bk_{DP} \right|, \label{electron_dispersion}
\end{equation} 
where $\eps$ is the band energy, $\hbar$ the reduced Planck constant,  $v_F$ the Fermi velocity, $\bf k$ the wave vector in the Brilloiun zone and $\bk_{DP}$ the position of $K$ Dirac point.  
In the homogeneous case, the electron dynamics in the semiclassical approximation is described by the Boltzmann equations

\begin{eqnarray}
	\dfrac{\partial f(t,\bk)}{\partial t} - \dfrac{e}{\hbar} \, E \, \frac{\partial f(t,\bk)}{\partial k_x} = \left. \dfrac{df}{d t}(t,\bk) \right|_{e-ph} \sv \label{bulk1}
\end{eqnarray}
where $f(t,\bk)$ is the electron distribution function at time $t$, ${\bk}=\left( k_x,k_y\right)$ the wave-vector, $e$ is the elementary (positive) charge and $E$ the applied external electric field which we take constant and along the $x$-axis. 

	The r.h.s. of Eq.~\eqref{bulk1} is the collision operator which describes the interactions of charge carriers with acoustic, optical and $K$ phonons. Acoustic phonons are intra-band, they preserve energy but not momentum and are longitudinal ($LA$) and transversal ($TA$); optical phonons, longitudinal ($LO$) and transversal ($TO$), can be also inter-band, i.e.~an electron can reach a band different from the one of the state before scattering, but in the same valley; $K$ phonons are optical and located near Dirac points, so their name, and can be also inter-valley, pushing the electron from a $K$ to a $K'$ valley. The general form of the collision operator is written as balance of gain and loss of charges in a given state ${\mathbf{k}}$:
\begin{eqnarray}
	&&
	\left. \dfrac{d f}{d t}(t,\bx,\bk) \right|_{e-ph} 
	= \left[
	\int_{{\cal B}} w(\bk', \bk) \,
	f(t,\bx,\bk') \left( 1 - f(t,\bx,\bk) \right) d \bk'  \right.
	\\
	&&
	\left. \hspace{92pt} -
	\int_{{\cal B}} w(\bk, \bk') \,
	f(t,\bx,\bk) \left( 1 - f(t,\bx,\bk') \right) d \bk' \right],
\end{eqnarray}
$w({\bf k'},{\bf k})$ is the total transition rate from a state $\bf k$ to a state $\bf k'$ and collects the contributions of collisions with each type of phonons and reads as

\begin{eqnarray}
	w(\bk', \bk)\! &=&\!
	\sum_{{\mu}} \left| G^{({\mu})}(\bk', \bk) \right|^{2} \!
	\left[  \left( g^{-}_{{\mu}} + 1 \right)
	\delta\! \left( \eps(\bk) - \eps(\bk') + \hbar \, \omega_{\mu}
	\right)  \nonumber \right. \\
	&& \left. + g^{+}_{{\mu}} \,
	\delta \left( \eps(\bk) - \eps(\bk') - \hbar \, \omega_{\mu}
	\right) \right], \quad   \label{Scatt}
\end{eqnarray}
where $\left| G^{({\mu})}(\bk', \bk) \right|^{2}$'s are the electron-phonon coupling matrix elements, $\mu= LA, TA, LO, TO, K$, $\delta$ is the Dirac distribution, $g_{\mu}^{\pm}=g_{\mu}({\bf q^{\pm}})$ are the phonon distributions, with wave vector $\bf q$ related to charge wave vector $\bf k$ by means of momentum conservation, i.e. ${\bf q}^{\pm}= \pm \left(\bk' - \bk \right)$.

The acoustic phonons transition in the elastic approximation \cite{Sarma} rate is given by

\begin{equation}
	w_{ac}(\bk', \bk)=\dfrac{1}{(2 \, \pi)^{2}} \, 
	\dfrac{\pi \, D_{ac}^{2} \, k_{B} \, T}{2 \hbar \, \sigma_m \, v_{\mu}^{2}}
	\left( 1 + \cos \vartheta_{\bk \sv \bk'} \right) \delta \left(\eps (\bk') - \eps (\bk) \right) , \quad \mu = LA, TA,
	\label{transport_acoustic}
\end{equation}
where $D_{ac}^{an}$ is the acoustic deformation potential, $\sigma_m$ is the density per unit area, $v_{\eta}$ the phonon branch sound speed 
and $\vartheta_{\bk \sv \bk'}$ is the angle forming final and in initial wave vectors. When the planar acoustic phonons are treated as a unique population, $w_{ac}^{is}$ has the same form of $w_{ac}^{an}$ with a single value $v_{ac}=2\times 10^4 \,\mbox{m/s}$ in place of $v_{LA}$ and $v_{TA}$: 

\begin{equation}
	w_{ac}^{is}(\bk', \bk)=\dfrac{1}{(2 \, \pi)^{2}} \, 
	\dfrac{\pi \, D_{ac}^{2} \, k_{B} \, T}{2 \hbar \, \sigma_m \, v_{ac}^{2}}
	\left( 1 + \cos \vartheta_{\bk \sv \bk'} \right) \delta \left(\eps (\bk') - \eps (\bk) \right)\,. 
	\label{transport_acoustic_is}
\end{equation}

The $LO$, $TO$, and $K$ coupling matrix elements present in Eq. (\ref{Scatt}) and which are used to calculate their contributions to the transition rate $w(\bk', \bk)$, when the planar optical population are treated as anisotropic, are given by \cite{EPC1}-\cite{EPC3}:

\begin{eqnarray}
	\left| G^{(LO)}_{an}(\bk', \bk) \right|^{2} & = & 
	\dfrac{1}{(2 \, \pi)^{2}} \, \dfrac{\pi \, D_{O}^{2}}{\sigma_m \, \omega_{O}}
	\left( 1 - \cos ( \vartheta_{\bk \sv \bk' - \bk} + \vartheta_{\bk' \sv \bk' - \bk} ) \right) \label{GLO}
	\\
	\left| G^{(TO)}_{an}(\bk', \bk) \right|^{2} & = & 
	\dfrac{1}{(2 \, \pi)^{2}} \, \dfrac{\pi \, D_{O}^{2}}{\sigma_m \, \omega_{O}}
	\left( 1 + \cos ( \vartheta_{\bk \sv \bk' - \bk} + \vartheta_{\bk' \sv \bk' - \bk} ) \right) \label{GTO}
	\\
	\left| G^{(K)}_{an}(\bk', \bk) \right|^{2} & = & 
	\dfrac{1}{(2 \, \pi)^{2}} \, \dfrac{2 \pi \, D_{K}^{2}}{\sigma_m \, \omega_{K}}
	\left( 1 - \cos \vartheta_{\bk \sv \bk'} \right) ,
\end{eqnarray}
where $D_{O}$ is the in plane optical phonon coupling constant, $\omega_{O}$ the in plane optical phonon frequency, $D_{K}$ is the $K$ phonon coupling constant and $\omega_{K}$ the $K$ phonon frequency; besides, $\vartheta_{\bk \sv \bk' - \bk}$ and $\vartheta_{\bk' \sv \bk' - \bk}$ are angles between $\bk$ and $(\bk' - \bk)$  and between $\bk'$ and  $(\bk' - \bk)$, respectively. In the isotropic approximation, we have to sum $\left| G^{(LO)}_{an}(\bk', \bk) \right|^{2}$ and $\left| G^{(TO)}_{an}(\bk', \bk) \right|^{2}$ and the angular dependence disappears

\begin{equation}
	\left| G^{(LO+TO)}_{is}(\bk', \bk) \right|^{2} = \dfrac{2}{(2 \, \pi)^{2}} \, \dfrac{\pi \, D_{O}^{2}}{\sigma_m \, \omega_{O}}. \label{GLO_is}
\end{equation}
We remark how isotropic approximation for planar acoustic phonons consists in considering a unique population with the same group velocity and the angular dependence in the determination of acoustic transition rate remains unaltered, while planar optical phonons become isotropic also geometrically.

Electron dynamics is a sequence of free flights in which semiclassical Hamilton's equations of motion hold

\begin{equation}
	\hbar \dot{\bf k}=-e{\bf E},
\end{equation}
interrupted by collision events with phonons. 

In the Direct Simulation Monte Carlo procedure (DSMC) the free flight duration $\delta t$ is determined by means of a random number $\xi$ uniformly distributed in $[0,1]$ as

\begin{equation}
	\label{free}
	\delta t=-\frac{\ln \xi}{\Gamma_{tot}},
\end{equation} 
where $\Gamma_{tot}$ is the total scattering rate, i.e. the probability per unit time that a scattering happens, and is given by the sum of the scattering rates of all the $\mu$th processes calculated starting from the transition rate $w_{\mu}$:
\begin{equation}
	\Gamma_{tot}({\bf k})=\sum_{\mu}\Gamma_{\mu}({\bf k})=\sum_{\mu}\int_{\cal B} w_{\mu}({\bf k'}, {\bf k}) d{\bf k'} \quad \mu= LA, TA, LO, TO, K.
\end{equation}
Out of plane $Z$-phonons are not included because they do not interact with electrons; they will be fundamental during phonon transport.

DSMC simulations consist of two main parts; in the first, the whole electron distribution $f(t,{\bf k})$ is rigidly translated in such a way that each charge carrier experiences the same free flight duration equal to the numerical time step $\Delta t=[t^{(j+1)}-t^{(j)}]$ and a change of momentum equal to

\begin{equation}
	\hbar \, \Delta  {\bf k}= -e\,{\bf E}\, \Delta t,
\end{equation} 
leading to an analytical solution of $f$, namely 

\begin{equation} \label{f_tr}
	f(t+\Delta t, {\bf k})=f(t,{\bf k}-\frac{e}{\hbar}{\bf E} \Delta t).
\end{equation}

We take as initial condition the Fermi-Dirac distribution at room temperature $T_R$
$$
f(0, \bk) = \dfrac{1}{1 + \exp \left( \dfrac{\eps(\bk) - \eps_F}{k_{B} \, T_R} \right)}.\\
$$

In the second part, starting from the electron distribution in Eq. \eqref{f_tr}, we consider individually the collisions of each particle
with phonons in a temporal window equal to the free flight duration the charge
carrier would have had if the global rigid translation had not been made, i.e.~$\delta t$. The procedure is repeated by calculating each $\delta t$ by means
of Eq.~\eqref{free} until $\Delta t$ is reached. The selection of the
scattering type and of the final state after collisions are based on the
generation of other random numbers uniformly distributed in $[0,1]$; we
refer the interested reader to~\cite{PRB}, in which the inclusion of Pauli's
exclusion principle is discussed as well.

Following the dynamics of collisions, we are able to count the number of emitted and absorbed phonons per unit time, $n_{\mu}^+({\mathbf{q}})$ and $n_{\mu}^-({\mathbf{q}})$, respectively, in each cell of the numerical grid, where the phonon wave vector ${\mathbf{q}}$ is deduced from the momentum conservation as mentioned above; they will enter the phonon dynamics as source terms.

Phonon transport in the homogeneous case is governed by the following Boltzmann equations for the distribution $g_{\mu}(t, {\mathbf{q}})$ of the $\mu$th phonon branch
\begin{equation}
	\frac{\partial g_{\mu}(t,{\mathbf{q}})}{\partial t} = C_{{\mu}-e}(t, {\mathbf{q}}) - \frac{g_{\mu}(t, {\mathbf{q}}) - g_{\mu}^{LE}(t, {\mathbf{q}})}{\tau_{\mu}\left(T_{\mu}\left(t\right)\right)}, \label{ph_tr}
\end{equation}
where
$C_{{\mu}-e}(t, {\mathbf{q}})$ describes the interactions between the $\mu$th phonons and
electrons and it is proportional to the net phonons production per unit time $\,$
$\left(n_{\mu}^{+}({\mathbf{q}})-n_{\mu}^-({\mathbf{q}})\right)$; the second term in the r.h.s. is the collision operator for the
phonon--phonon scatterings written in the Bhatnagar--Gross--Krook (BGK)
approximation, which models the tendency of each phonon population with
temperature $T_{\mu}$ to reach an equilibrium distribution locally
determined by a Bose--Einstein $g_{\mu}^{LE}$ evaluated at the crystal lattice
local equilibrium temperature $T_{LE}$. $\tau_{\mu}(T_{\mu})$ are the temperature
dependent relaxation times which from the frequency-integrated model
in~\cite{Bao,Hao,Peraud} can be written as
\begin{equation}
	\tau_{\mu} (T_{\mu}) = \sum_{k= 0}^{7} c_k \left(\frac{T_{\mu}}{T_{R}}\right)^k, \label{tau2}
\end{equation}
where the values of $c_k$ are reported in~\cite{CR_HT}.

Given the temporal grid $[t^{(j)}, \,\, t^{(j)} + \Delta t]$, with an uniform time step $\Delta t$, we solve Eq.~\eqref{ph_tr} in each $\nu$th cell of the numerical Brillouin zone with wave vector ${{\mathbf{q}}^{\nu}}=(q^{\nu}_x, q^{\nu}_y)$ by using an explicit Euler method, being the DSMC results already affected by a numerical error of the first order in $\Delta t$, as follows
\begin{equation}
	g_{\mu}^{(j+1)}=g_{\mu}^{(j)}+ \Delta t \, \frac{\alpha_v\beta_{el}}{\left| C_{\nu}\right| } \left(n_{\mu}^{+}({{\mathbf{q}}^{\nu}})-n_{\mu}^{-}({{\mathbf{q}}^{\nu}}) \right)^{(j)}- \Delta t \, \frac{g_{\mu}^{(j)} - g_{\mu}^{LE (j)}}{\tau_{\mu}\left(T_{\mu}\left(t^{(j)}\right)\right)}\,, \label{num_sol_ph}
\end{equation}
where $\alpha_v=2$ is the valley degeneracy, $\beta_{el}=\rho/N_P$ the electron statistical weight, being $\rho$ the graphene (areal) electron density and $N_P$ the number of simulated (super)-particles,  and $\left| C_{\nu}\right| $ the measure of the $\nu$th cell. As initial condition we take the Bose--Einstein distribution
\[
g_{\mu}(0, {\mathbf{q}}) = \dfrac{1}{-1 + \exp \left( \dfrac{\hbar \omega_{\mu}}{k_{B} \, T_R} \right)}.
\]

To be able to perform the computation in Eq.~\eqref{num_sol_ph}, we need to know the temperatures of each phonon population $T_{\mu}(t^{(j)})$ and that of the crystal lattice local equilibrium $T_{LE}(t^{(j)})$. At time $t^{(j)}$, from $g_{\mu}(t^{(j)}, {\mathbf{q}})$, we know the numerical value of the energy density of each $\mu$th phonon branch, $W_{\mu}$, which is locally taken as equal to that given by an equivalent Bose--Einstein distribution at temperature $T_{\mu}$:
\begin{equation}
	W_{\mu}= \frac{1}{(2 \pi)^2} \int_{\mathcal{B}}  \hbar \omega_{\mu} g_{\mu}(t, {\mathbf{q}})
	\, d {\mathbf{q}} = \frac{1}{(2 \pi)^2} \int_{\mathcal{B}}  \hbar \omega_{\mu} \, \left[\displaystyle{e^{\hbar \omega_{\mu}/ k_B T_{\mu}}}  - 1\right]^{-1}  
	\, d {\mathbf{q}}. \label{num_mu}
\end{equation}

$T_{\mu}$ is given by numerical inversion of Eq.~\eqref{num_mu}. $T_{LE}$ is calculated by exploiting the following property of the phonon collision operator
\begin{equation}
	\sum_{\mu} \frac{W_{\mu} - W_{\mu}^{LE}}{\tau_{\mu}(T_{\mu})}  = 0 , \label{prod_ph}
\end{equation}
where
\begin{equation}
	  W_{\mu}^{LE} = \frac{1}{(2 \pi)^2} \int_{\mathcal{B}} \hbar \omega_{\mu} \, \left[\displaystyle{e^{\hbar \omega_{\mu}/k_B T_{LE}} } - 1\right]^{-1} \label{W_eq_1}
	\, d {\mathbf{q}},
\end{equation}
and $W_{\mu}$'s are given by Eq.~\eqref{num_mu}. 

The alternative definition of $T_{LE}$ is obtained by replacing Eq.~\eqref{prod_ph} with the following
\begin{equation}
	\sum_{\mu} W_{\mu}=\sum_{\mu} W_{\mu}^{LE} \quad \mu=LA, TA, ZA, LO, TO, ZO, K, \label{second_T}
\end{equation}
in which relaxation times do not explicitly enter.

For $LA$ and $TA$ phonons we use the Debye approximation $\omega_{\mu} = v_{\mu} q $, $\mu = LA, TA$, while for $ZA$ phonons we have $\omega_{ZA} = \gamma \left| {\bf q}\right| ^{2} $, with $\gamma = 6.2 \cdot 10^{-7}$m$^{2}$/s (\cite{MoMa,NiBa,Pop}). For optical
phonons, the Einstein approximation  of constant $\hbar \omega_{LO, TO, K}$'s is adopted. We
refer to~\cite{CR_HT} and \cite{Coco_IJMPB} for further details.

\section{Numerical results}
\label{results}

\begin{table}[h] 
	\caption{Physical parameters used in the simulations.\label{table1}}
	\centering
		\begin{tabular}{|c|c|}
			\hline
			$\sigma_m$ & $7.6 \times 10^{-8}$ g/cm$^2$\\
			$v_F$         & $10^6$ m/s\\ 
			$v_{LA}$       & $ 2.13 \times 10^4$ m/s\\
			$v_{TA}$       & $ 1.36 \times 10^4$ m/s\\
			$D_{ac}$    & $6.8$ eV \\ 
			$ \hbar \, \omega_{LO}$ & $164.6$ meV\\
			$ \hbar \, \omega_{TO}$ & $164.6$ meV\\
			$D_{O}$    & $10^9$ eV/cm\\ 
			$ \hbar \,\omega_{K}$ & $124$ meV \\
			$D_{K}$ & $3.5 \times 10^8$ eV/cm\\
			\hline
		\end{tabular}
\end{table}

	In the simulations, we use $N_P=10^4$ (super)-particles and the time step is
$\Delta t=2.5$ fs; the electrons Brillouin zone is discretized by a uniform
square grid $[-k_{x max}, k_{x max}]\times[-k_{y max}, k_{y max}]$ centered in the $K$ point, with $k_{x max}=k_{y max}=24 \, \mbox{nm}^{-1}$,
and {1042~$\times$~1042} cells; for the phonons space we use a square grid with
322~$\times$~322 cells; the $\mathbf{q}$-vector space grid is centered at the
center $\Gamma$ of the Brillouin zone. The choice of the centers is done in
such a way to not introduce any numerical error by substituting the hexagonal
BZ with a square grid because the charge and phonon distributions are different
from zero only around the $K$ and $\Gamma$ point, respectively.
For the physical parameters, we refer to the values presented
in~\cite{Li2010,Bor}, and reported in 
\ref{table1}. We consider in the following the case of a Fermi level $\varepsilon_F=0.6~{\rm eV}$ and an applied electric field $E=20$ kV/cm. 

We distinguish four main situations: when all the planar phonons are isotropic, when only the acoustic or the optical are anisotropic, acoustic phonon anisotropy (APA) and optical phonon anisotropy (OPA), respectively, and the case in which all the in-plane phonon branches are anisotropic.

We take at first the definition of $T_{LE}$ given by Eq.~\eqref{prod_ph}.

\begin{figure}[h!]
	\centering
	\includegraphics[width=0.6\columnwidth]{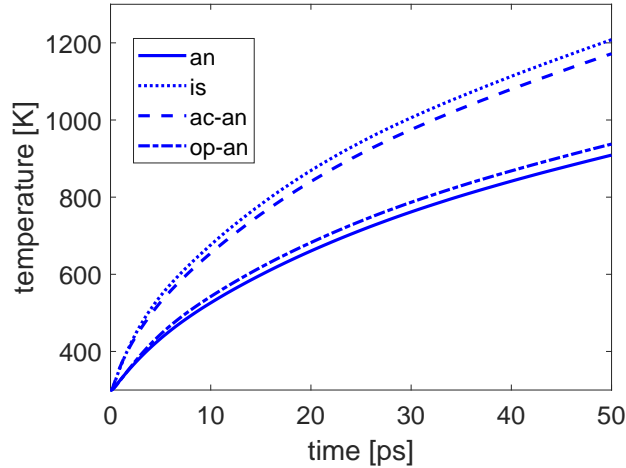}\\
	\caption{Local equilibrium temperature $T_{LE}$ with (an) and without (is) anisotropy, and when only acoustic (ac-an) or optical (op-an) phonon anisotropy is considered. $\varepsilon_F=0.6~{\rm eV}$ and $E=20$ kV/cm.	\label{T_LE}}
\end{figure}

As shown in  \ref{T_LE}, the inclusion of planar phonon anisotropy affects in a remarkable way the evaluation of the local equilibrium temperature $T_{LE}$. In the isotropic case, it is higher of about 33\% with respect to the anisotropic one and heating effects are definitely overestimated. The optical phonon anisotropy (OPA) is more effective than the acoustic phonon anisotropy (APA): inclusion of only OPA produces a slight increasing of the anisotropic temperature while only APA is not able to appreciably reduce the temperature with respect to isotropic approximation.

\begin{figure}[h!]
	\centering
	\fbox {a)		\includegraphics[width=0.41\columnwidth]{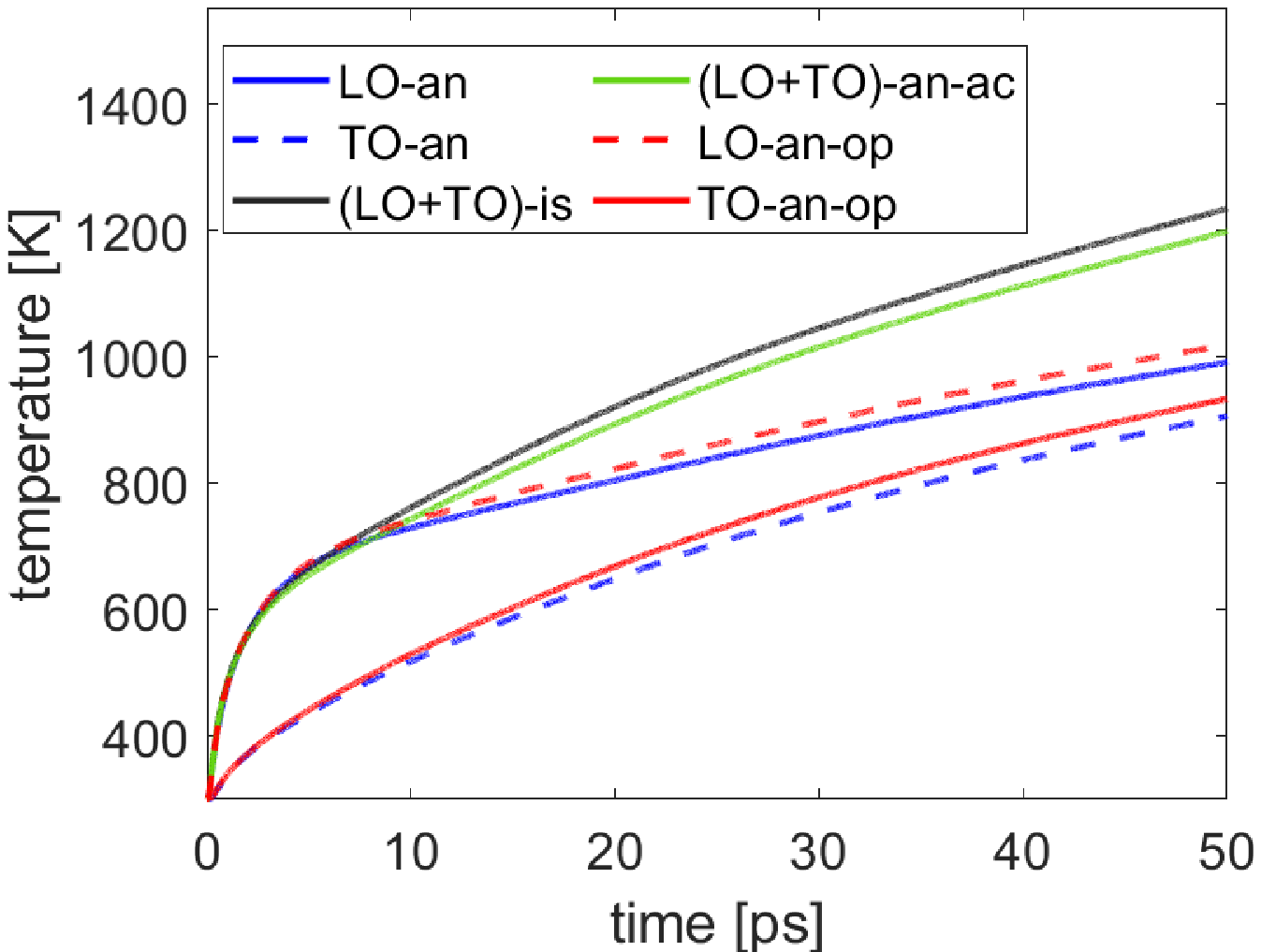}}
	\fbox {b)		\includegraphics[width=0.41\columnwidth]{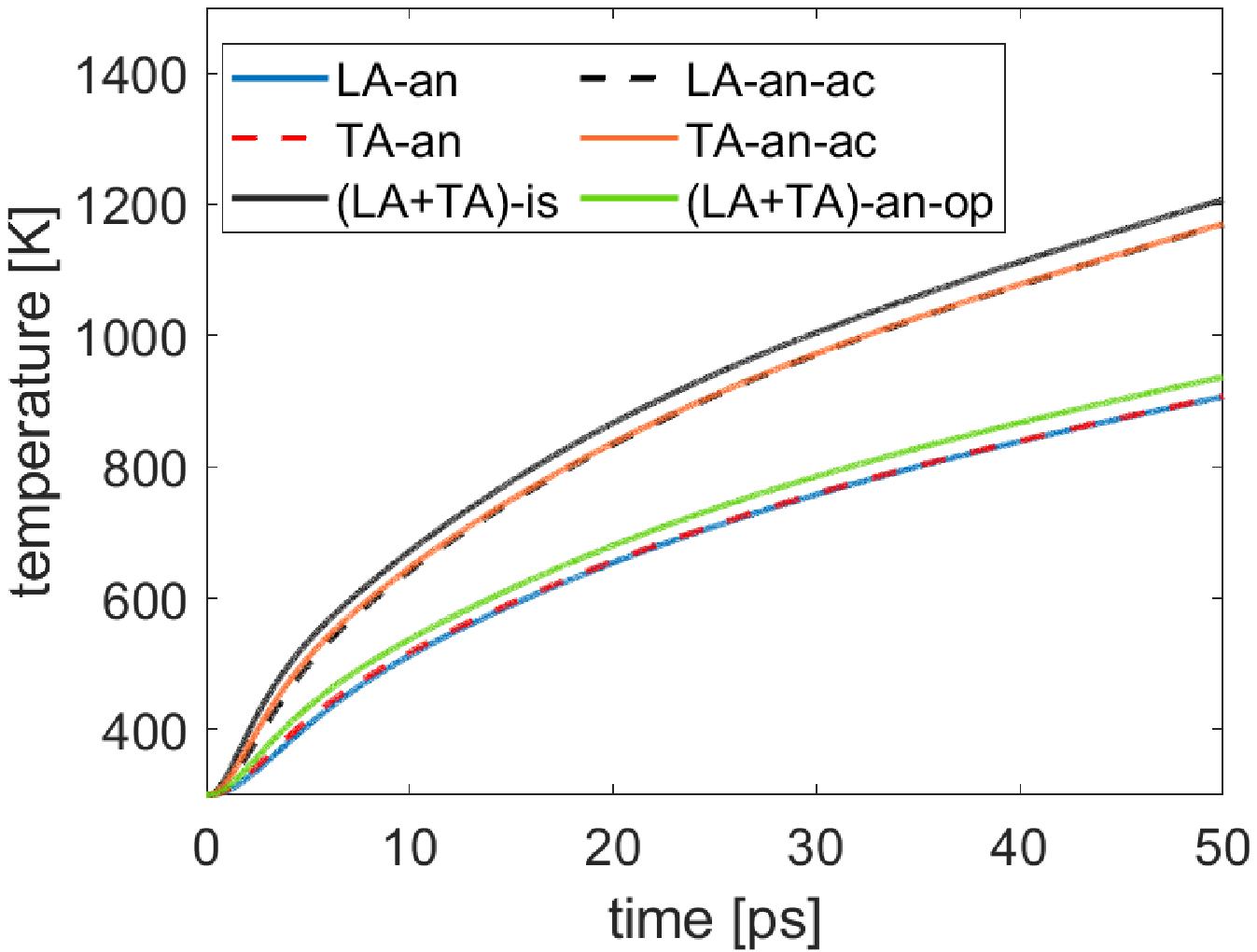}}\\
	\fbox {c)		\includegraphics[width=0.41\columnwidth]{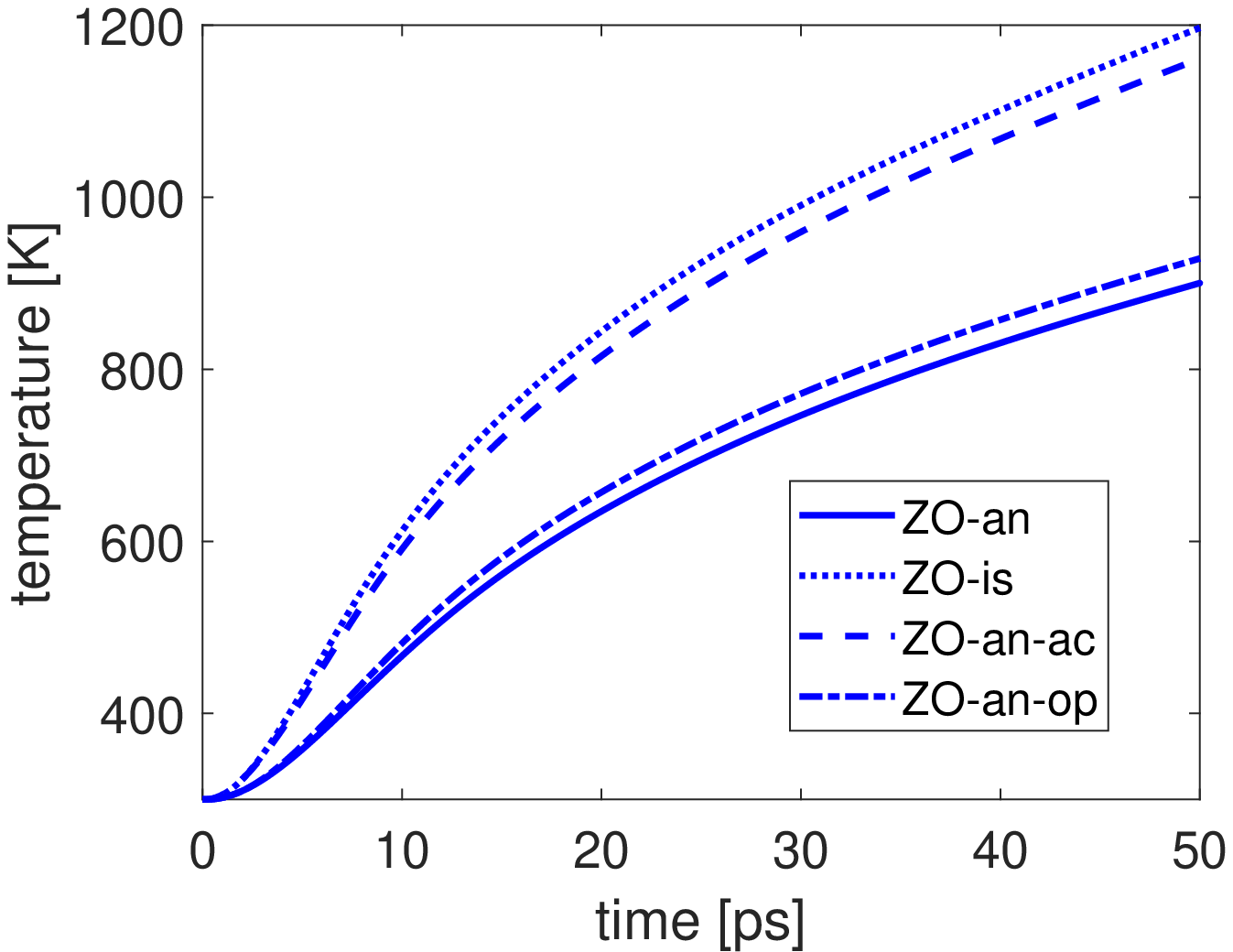}}
	\fbox {d)		\includegraphics[width=0.41\columnwidth]{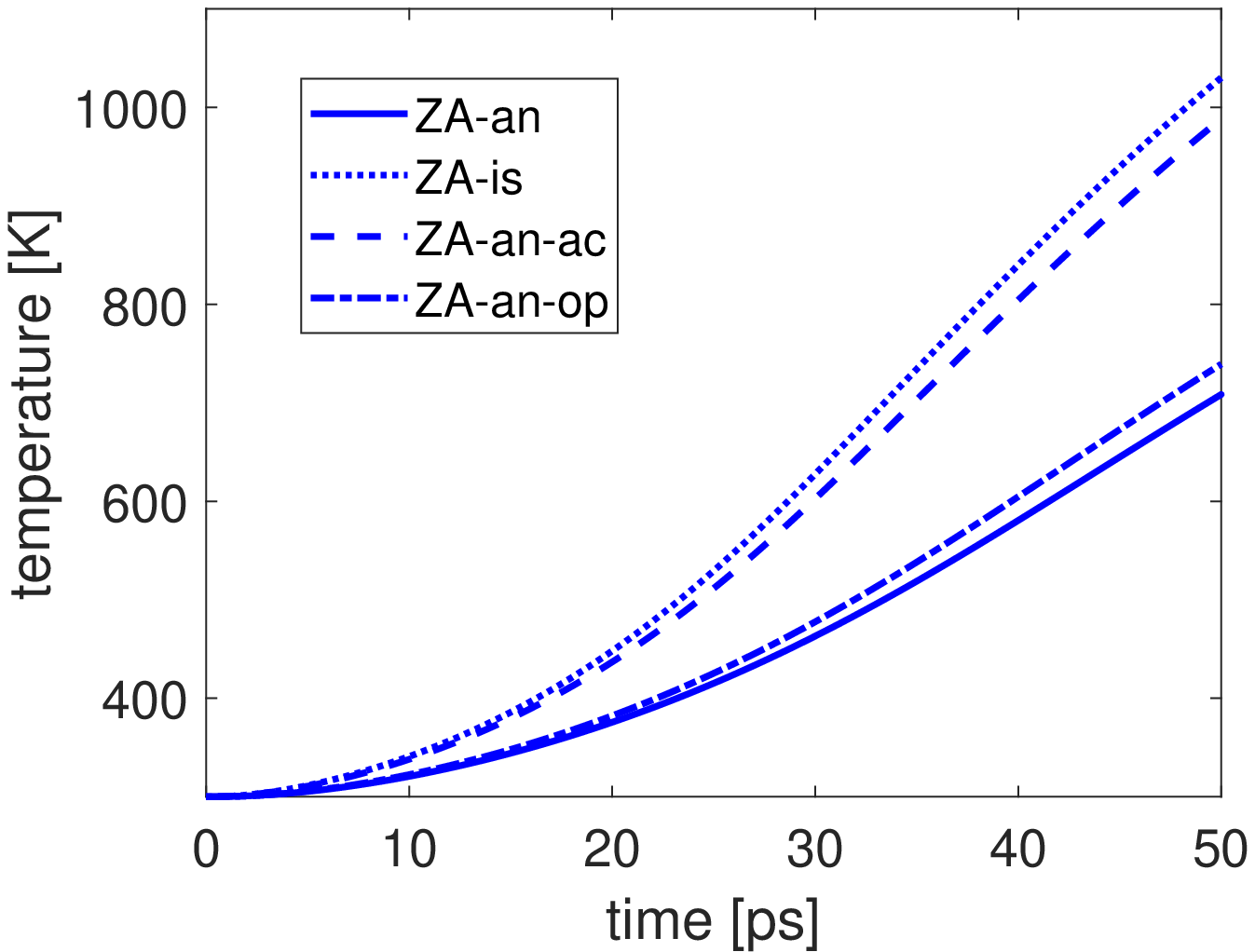}}\\		
	\caption{Planar phonon temperatures with (an) and without (is) anisotropy, and when only acoustic (an-ac) or optical (op-an) phonon anisotropy is considered. $\varepsilon_F=0.6~{\rm eV}$ and $E=20$ kV/cm.	\label{T_pl}}
\end{figure}

\begin{figure}[h!]
	\centering
	\includegraphics[width=0.6\columnwidth]{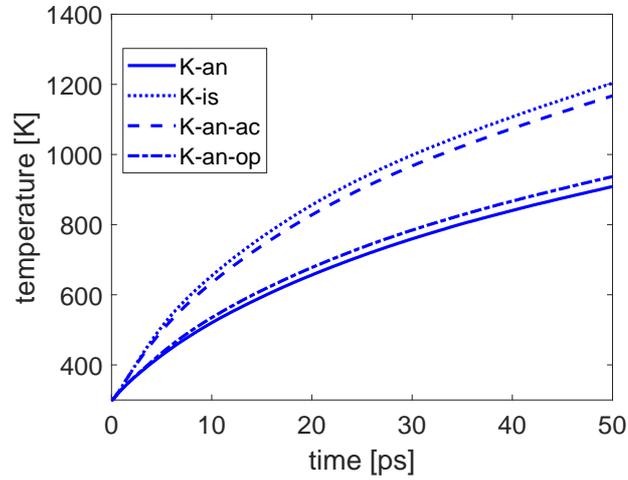}\\
	\caption{$K$ phonon temperature with (an) and without (is) anisotropy, and when only acoustic (an-ac) or optical (op-an) phonon anisotropy is considered. $\varepsilon_F=0.6~{\rm eV}$ and $E=20$ kV/cm.	\label{T_K}}
\end{figure}

Inclusion of anisotropy produces a great difference also in the values of the temperature of each phonon branch (Fig.s \ref{T_pl}--\ref{T_K}), reaching a discrepancy even of about 40\% for the $ZA$ phonons (Fig. \ref{T_pl} panel d), which confirm their fundamental role in thermal study of graphene. APA is negligible also in the evaluation of each phonon branch temperature, producing only a small change in the results. 

\begin{figure}[h!]
	\centering
	\includegraphics[width=0.6\columnwidth]{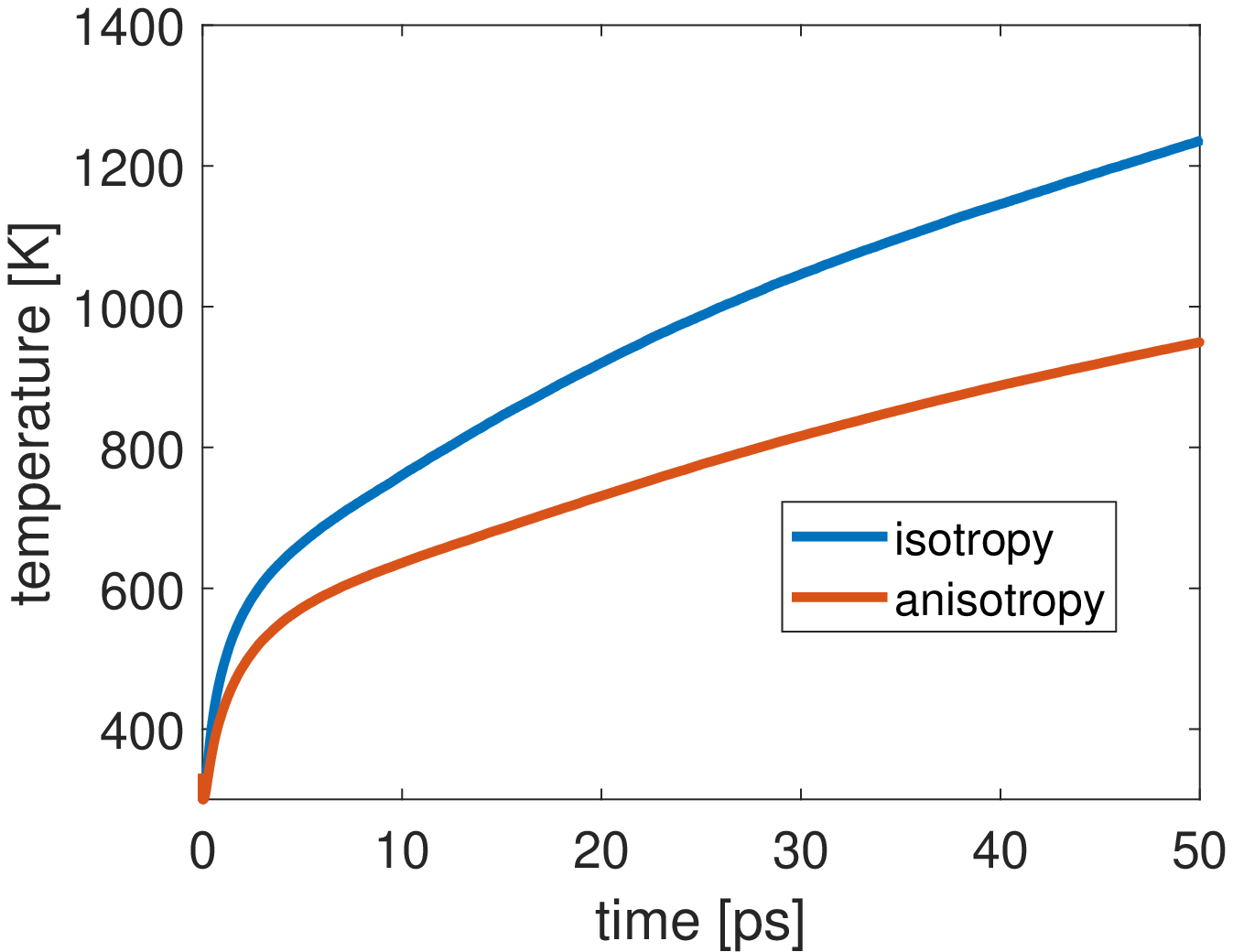}\\
	\caption{Temperature of planar optical branch $(LO+TO)$ in the isotropic case (blue line) and equivalent $T_{op}^{an}$ in the anisotropic case (red line). $\varepsilon_F=0.6~{\rm eV}$ and $E=20$ kV/cm.	\label{T_op_eq}}
\end{figure}

	To highlight in particular the temperature evolution of the optical phonons, which are the most relevant during the simulation, in  \ref{T_op_eq} we compare the temperature of the planar optical phonons $(LO+TO)$ in the isotropic case and the equivalent temperature, $T_{op}^{an}$, $LO$ and $TO$ populations would have in the anisotropic case if they were considered as a unique population, calculated by inverting the following relation
\begin{equation}
	W_{LO+TO}^{an}=\frac{2}{(2 \pi)^2}\int_{\mathcal{B}} \frac{\hbar \omega_{op}}{-1+\exp{\frac{\hbar \omega_{op}}{k_B T_{op}^{an}}}} d{\mathbf{q}}= W_{LO}^{an}+W_{TO}^{an},
\end{equation}
where $W_{LO}^{an}$ and $W_{TO}^{an}$ are given by Eq.~\eqref{num_mu}. We remind that $\hbar\omega_{op}$ is equal for $LO$ and $TO$ phonons. The optical phonon temperature in the isotropic case reaches a value higher of about 40\%.

\begin{figure}[h!]
	\centering
	\fbox {a)		\includegraphics[width=0.41\columnwidth]{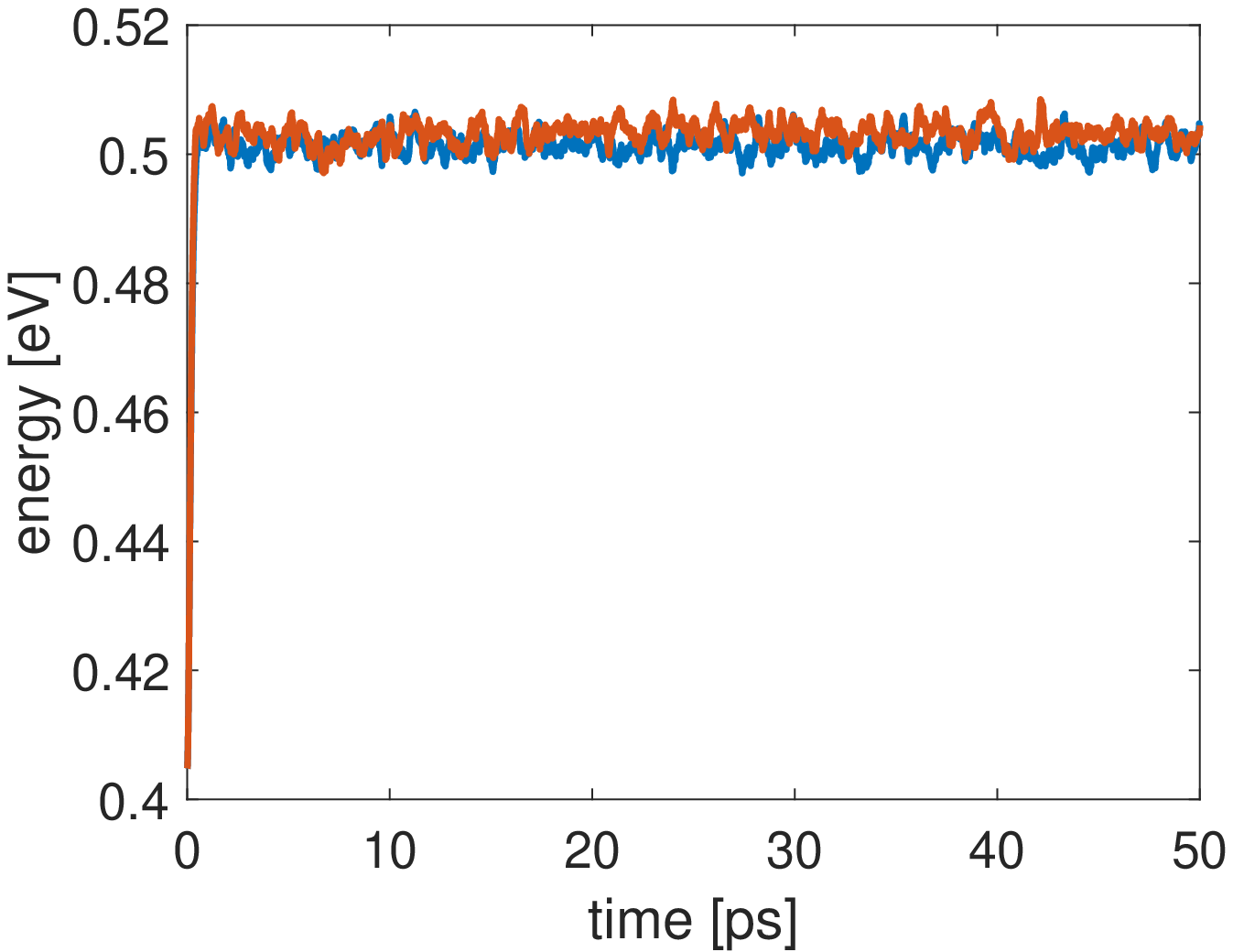}}
	\fbox {b)		\includegraphics[width=0.41\columnwidth]{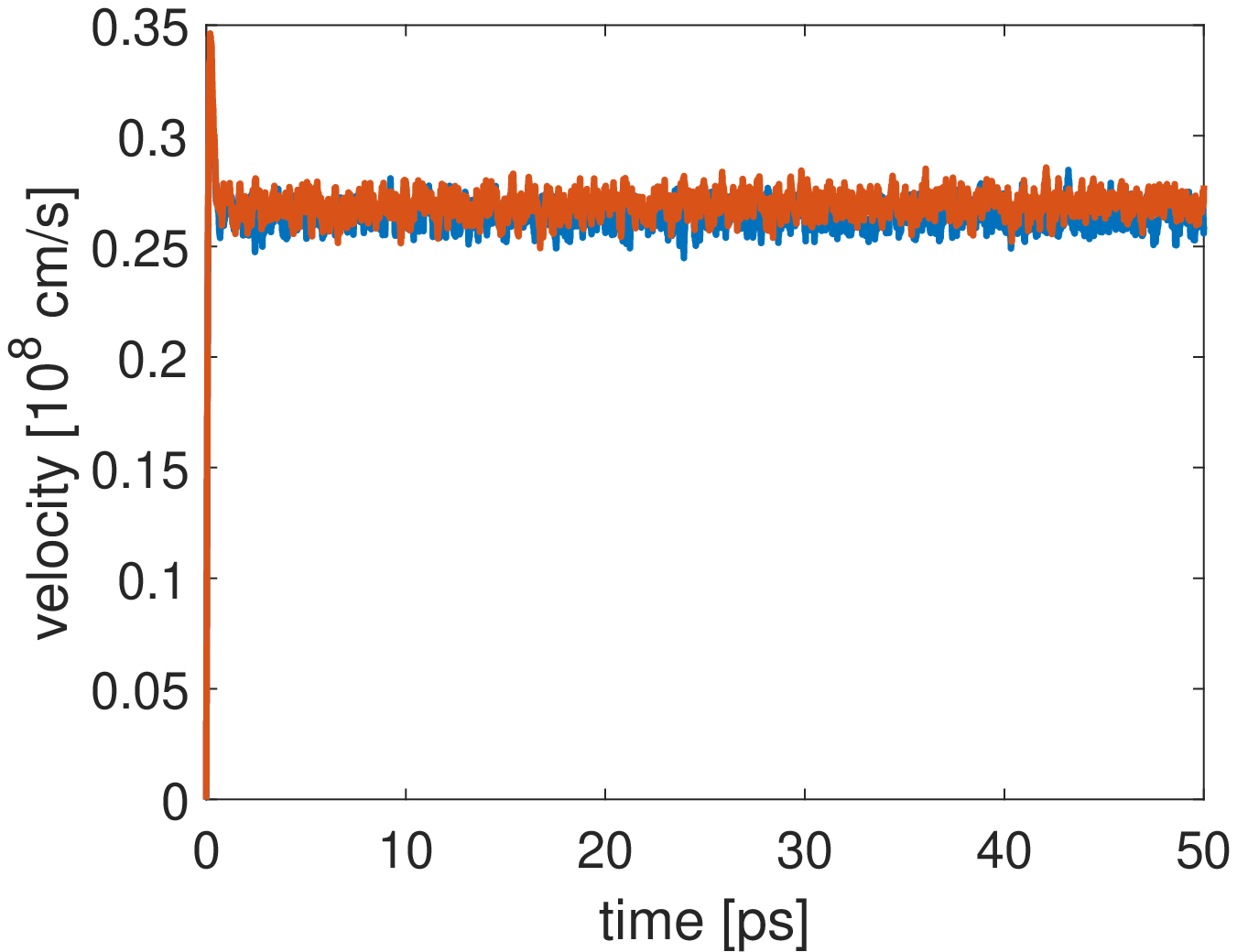}}\\
	\caption{Mean energy, (a), and velocity, (b), with (blue line) and without anisotropy (red line), when phonon are kept at thermal bath. $\varepsilon_F=0.6~{\rm eV}$ and $E=20$ kV/cm. (For
		interpretation of the references to colors in this figure, the reader is referred to the web version of this article.)	\label{el_noT}}
\end{figure}

\begin{figure}[h!]
	\centering
	\fbox {a)		\includegraphics[width=0.41\columnwidth]{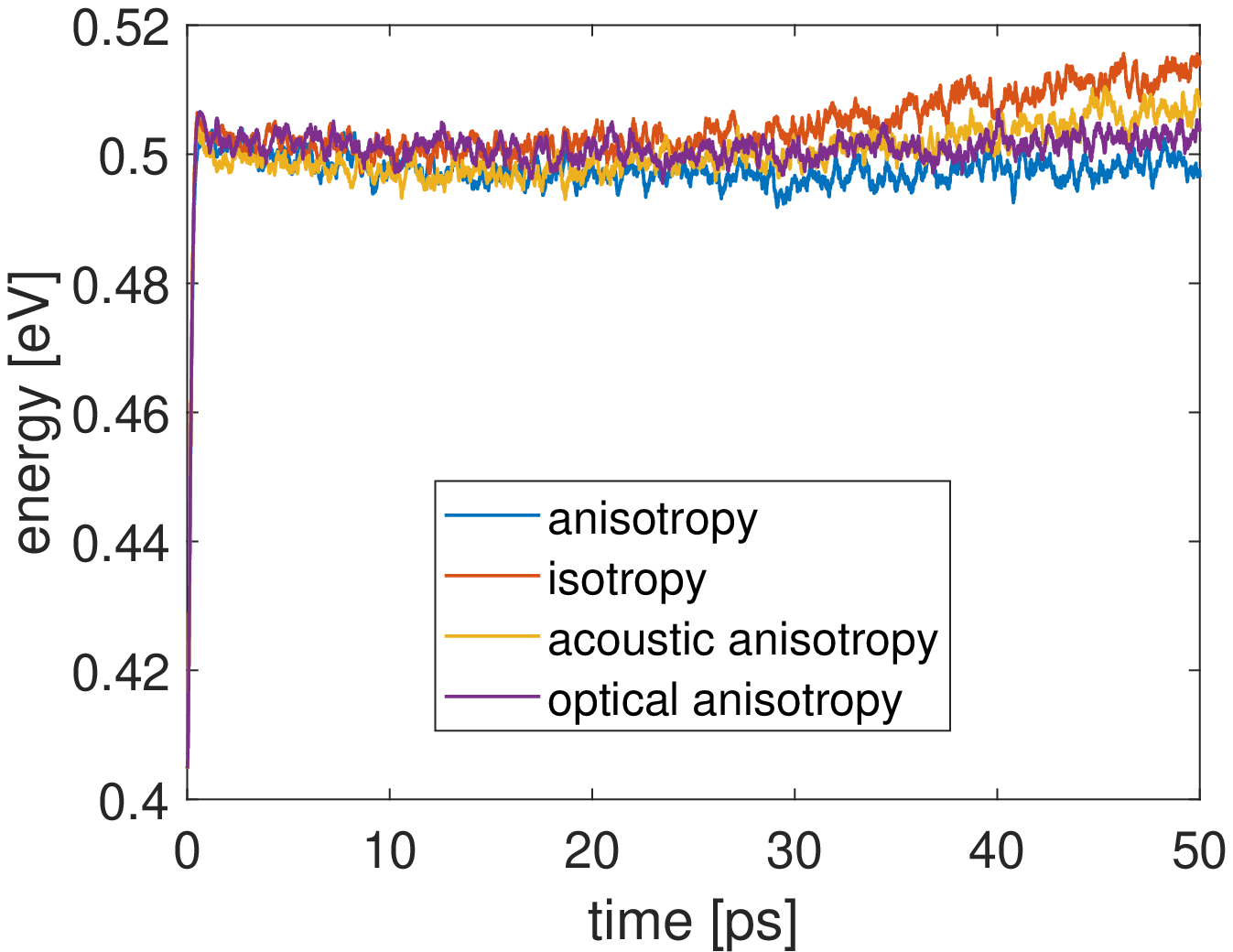}}
	\fbox {b)		\includegraphics[width=0.41\columnwidth]{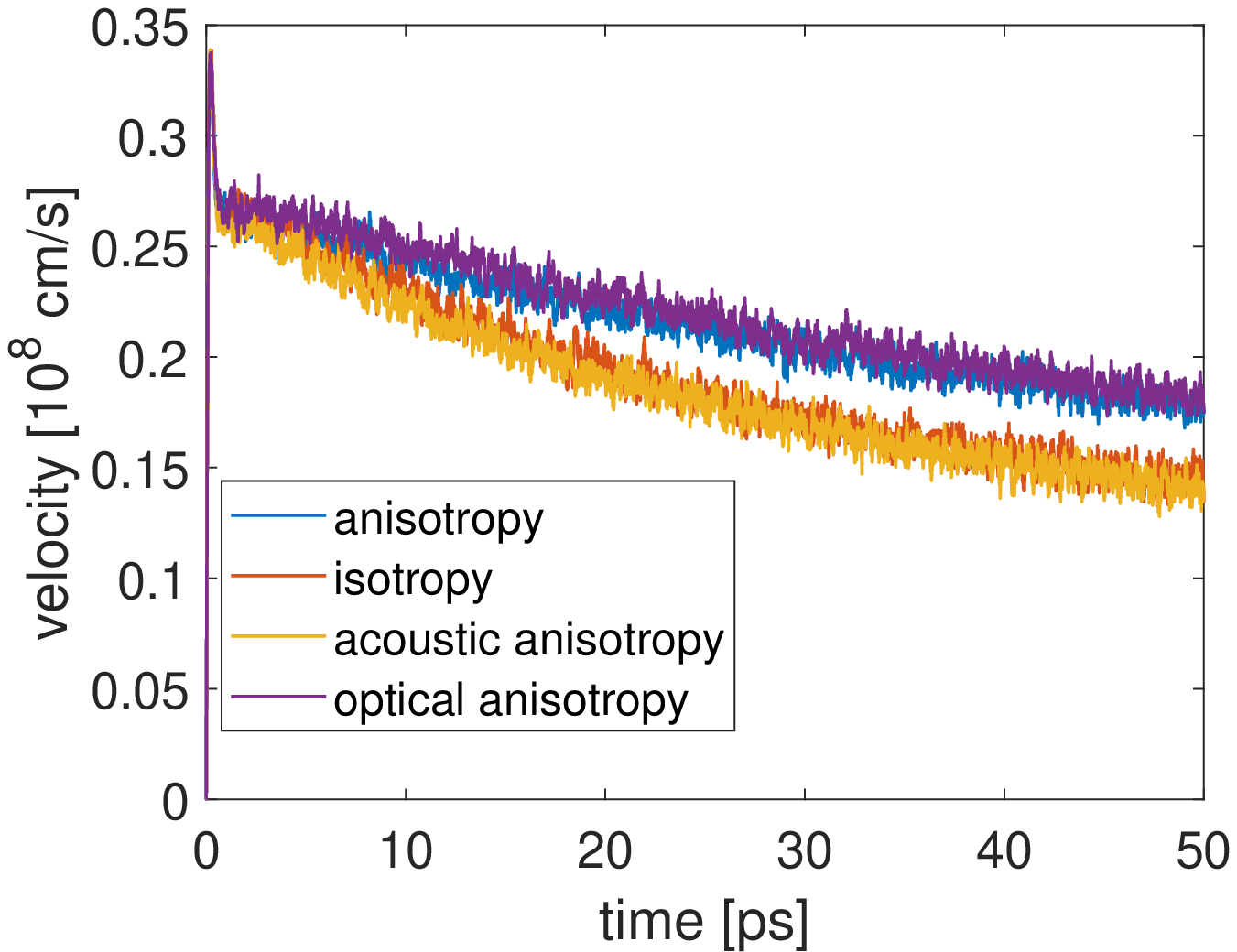}}\\
	\caption{Mean energy, (a), and velocity, (b), with and without anisotropy, and when only acoustic or optical phonon anisotropy is considered. $\varepsilon_F=0.6~{\rm eV}$ and $E=20$ kV/cm. (For
		interpretation of the references to colors in this figure legend, the reader is referred to the web version of this article.)	\label{el_T}}
\end{figure}

	The different evaluation of local equilibrium temperature due to anisotropy affects the graphene electric characteristic curves as well. In Fig. \ref{el_noT}  the charge mean energy and velocity are shown with and without phonon anisotropy but when phonons are kept at thermal bath; in this case, any effect of anisotropy is present. When hot phonon transport is included, the situation results rather different as Fig.  \ref{el_T} shows. In the left panel, the mean energy reveals an unexpected behavior: in the anisotropic case, it reaches a steady state value (blue line), while in all the other cases it increases after $20$ ps; the lower the anisotropy contribution taken into account is, the higher the final values are; a slight variation is present when only the optical phonon anisotropy is included while is higher with only the acoustic phonon anisotropy, and the isotropic case reaches at $50$ ps a difference of about 3\% with respect to the anisotropic value. In the right panel of Fig. \ref{el_T}, the electron mean velocity is reported; it shows the typical degradation due to heating effects which is definitely more relevant in the isotropic case, with a value lower with respect to the anisotropic one of about 16\% at $50$ ps. Moreover, for average velocity, OPA and APA do not introduce any relevant contribution and the values obtained with only optical or acoustic anisotropy are superposed to anisotropic and isotropic case, respectively.
	
In the following, we will discuss in detail how neglecting the planar phonons anisotropy affects the physical processes that occur during simulations and we will explain the previous results as well. We will take particular advantage from the typical possibility given by Monte Carlo procedure to look into each single part of the physical simulated phenomenon.   

\begin{figure}[h!]
	\centering
	\includegraphics[width=0.6\columnwidth]{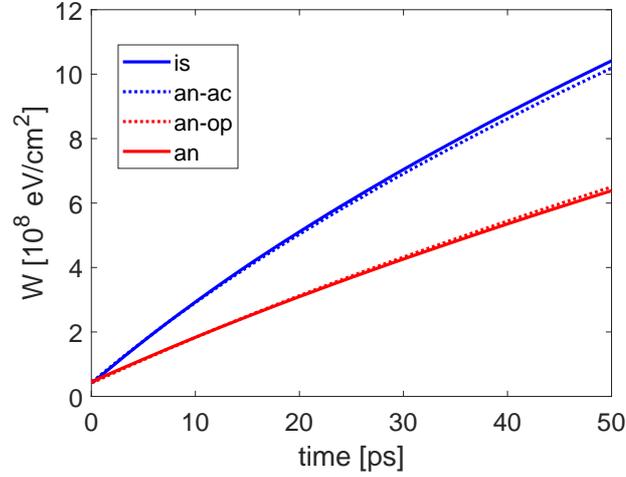}\\
	\caption{Total phonon energy density with (an) and without (is) anisotropy, and when only acoustic (an-ac) or optical (an-op) phonon anisotropy is considered. $\varepsilon_F=0.6~{\rm eV}$ and $E=20$ kV/cm.	\label{W_tot}}
\end{figure}

	As first step, we analyze the temporal evolution of the phonon energy density, which is directly related to the temperature evolution by means of Eqs.~\eqref{num_mu}--\eqref{prod_ph}, where also the presence of temperature dependent relaxation times has to be considered.  In Fig.  \ref{W_tot}, we report the total energy density, sum of that of each single phonon branch; also the energy density, as the local equilibrium temperature, presents a great difference between the isotropic and not isotropic case with a gap of about 66\% at $50$ ps, and the isotropic approximation leads to consider a really higher energetic content in the study of thermo-electrical behavior of graphene. Also for the phonon energy density values, the APA and OPA reveal a negligible contribution. 
 
\begin{figure}[h!]
	\centering
	\fbox {a)		\includegraphics[width=0.41\columnwidth]{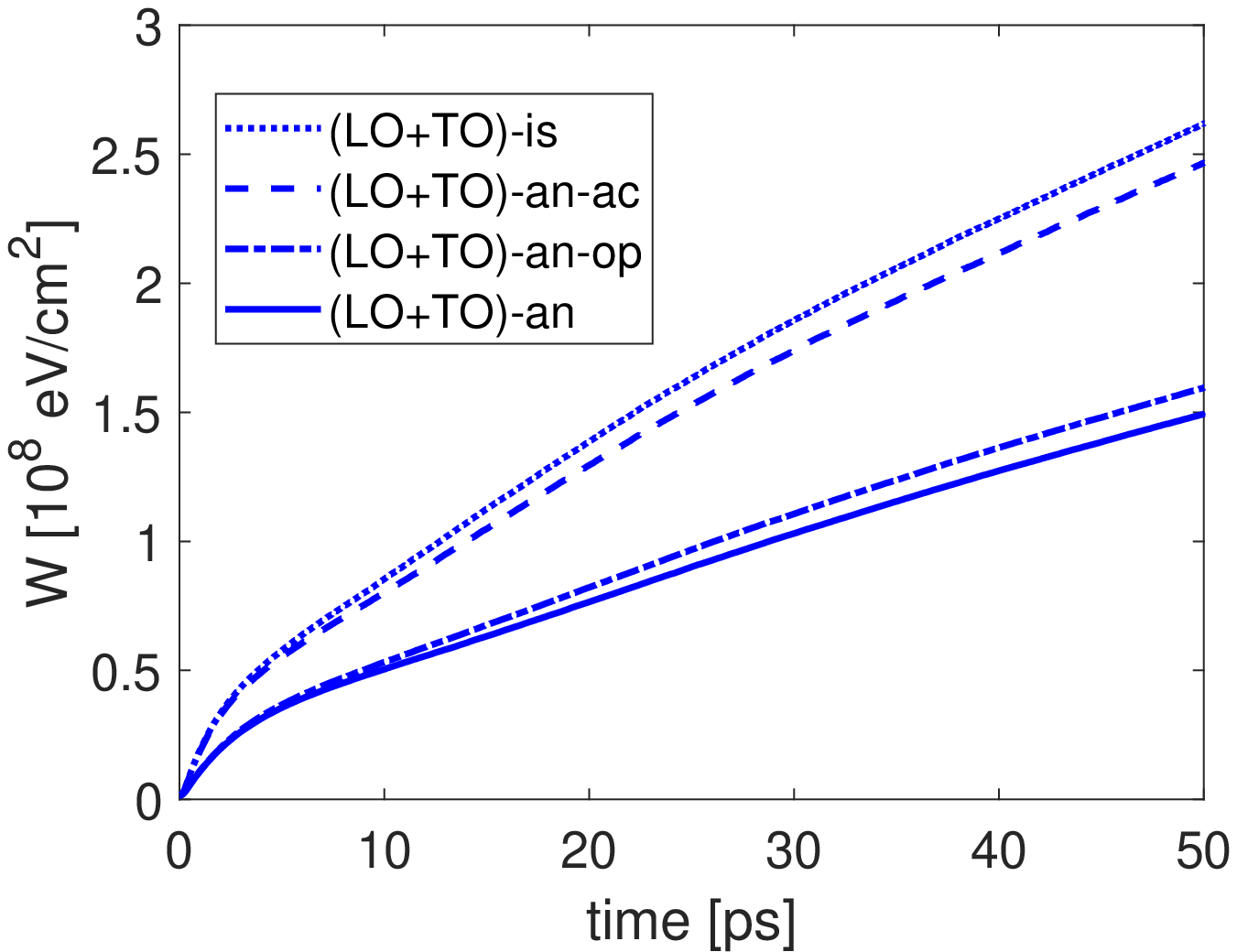}}
	\fbox {b)		\includegraphics[width=0.41\columnwidth]{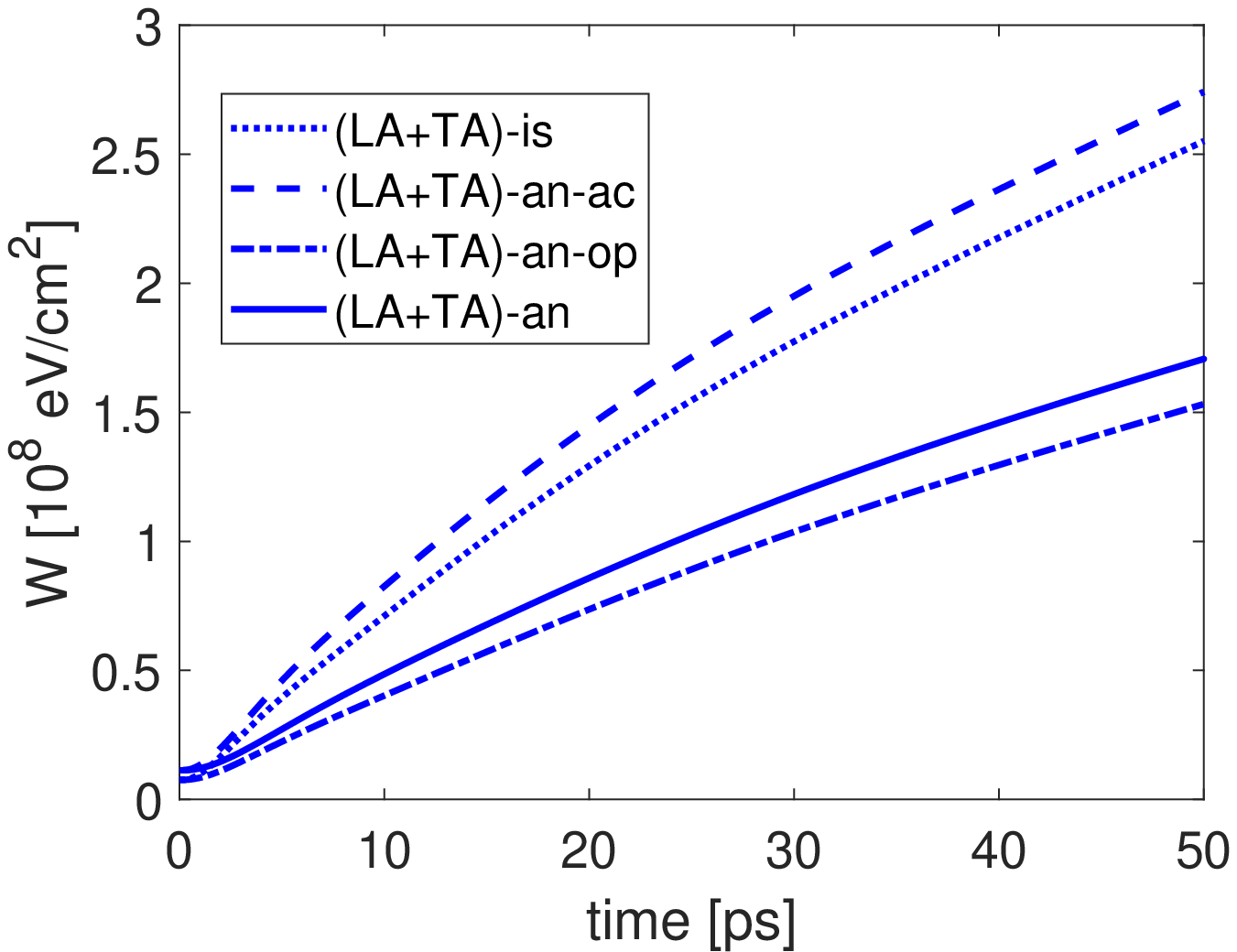}}\\
	\fbox {c)		\includegraphics[width=0.41\columnwidth]{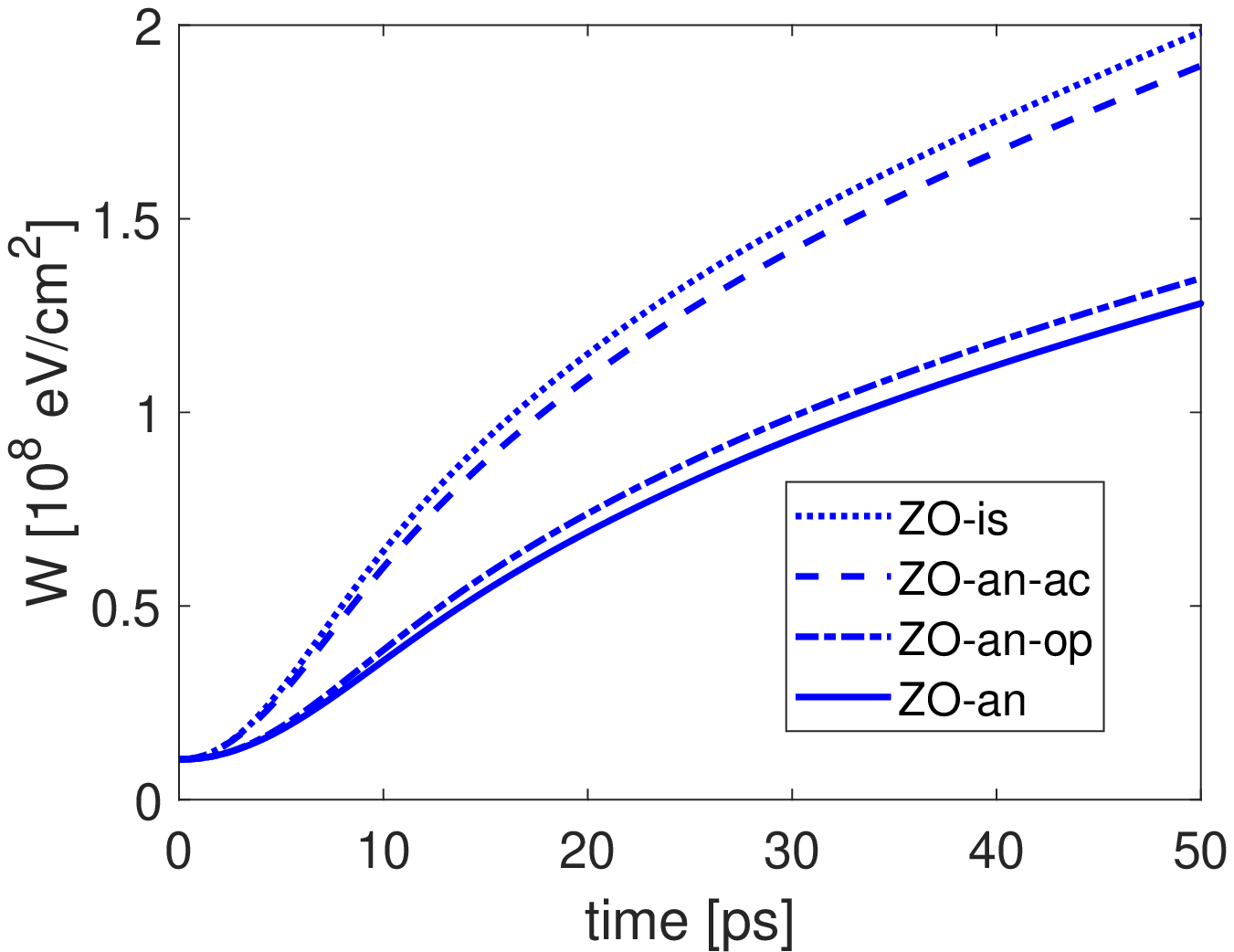}}
	\fbox {d)  \includegraphics[width=0.41\columnwidth]{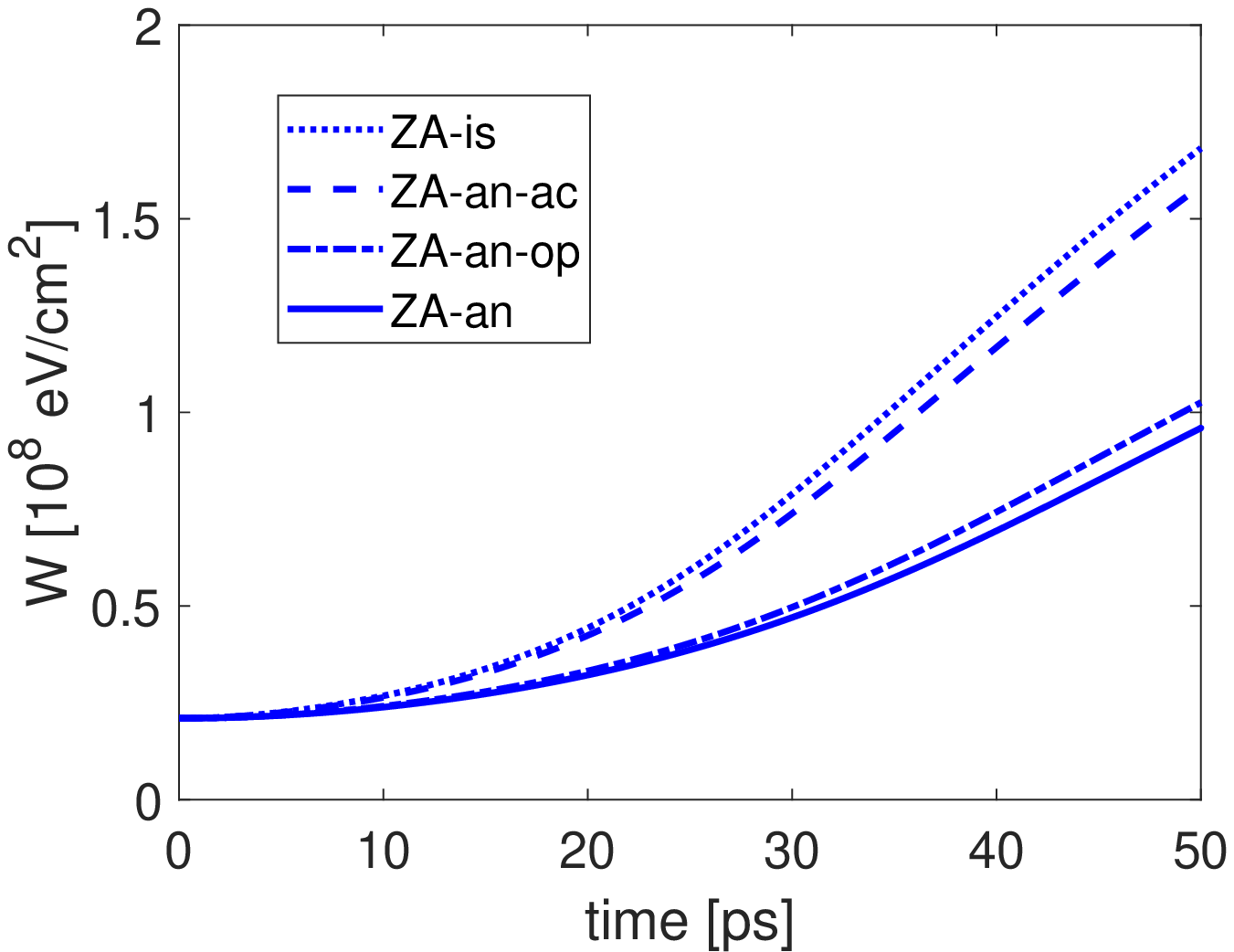}}\\		
	\caption{Planar optical, acoustic, $ZO$ and $ZA$ phonon energy densities, with (an) and without (is) anisotropy, and when only acoustic (an-ac) or optical (an-op) phonon anisotropy is considered. $\varepsilon_F=0.6~{\rm eV}$ and $E=20$ kV/cm.	\label{W_ph1}}
\end{figure}

\begin{figure}[h!]
	\centering
	\includegraphics[width=0.6\columnwidth]{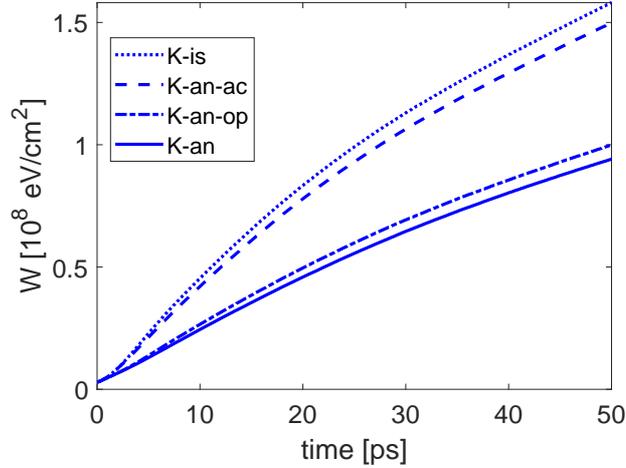}\\
	\caption{$K$ phonon energy density, with (an) and without (is) anisotropy, and when only acoustic (an-ac) or optical (an-op) phonon anisotropy is considered. $\varepsilon_F=0.6~{\rm eV}$ and $E=20$ kV/cm.	\label{W_K}}
\end{figure}

This is confirmed for the energy density of each phonon population as well, as shown in Fig.s \ref{W_ph1}--\ref{W_K}. The curves obtained in the isotropic and anisotropic case are the highest and the lowest one, respectively, except for the in-plane acoustic phonons in Fig. \ref{W_ph1}(b), in which with only OPA the energy density decreases and with only APA the energy density increases with respect to the anisotropic and isotropic curves, respectively. In Fig.  \ref{W_ph1}(a) and (b), we have taken into account a cumulative contribution of the planar phonons, i.e.~equivalent $(LA+TA)$ and $(LO+TO)$ populations, also in the anisotropic cases for the sake of comparison with the isotropic results. The behavior of the phonon energy density is coherent with the temperature evolution.

\begin{figure}[h!]
	\centering
	\fbox {a)		\includegraphics[width=0.41\columnwidth]{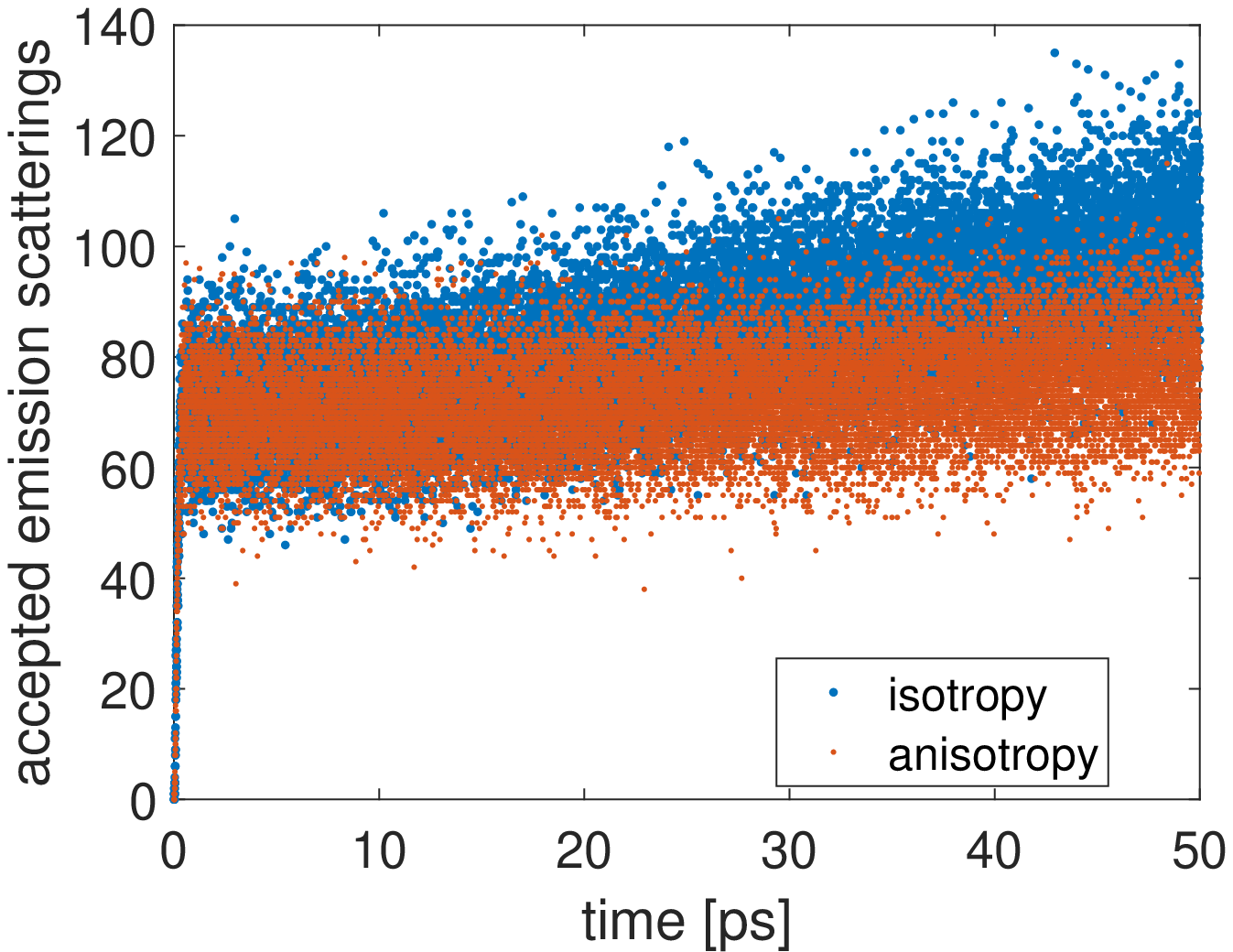}}
	\fbox {b)		\includegraphics[width=0.41\columnwidth]{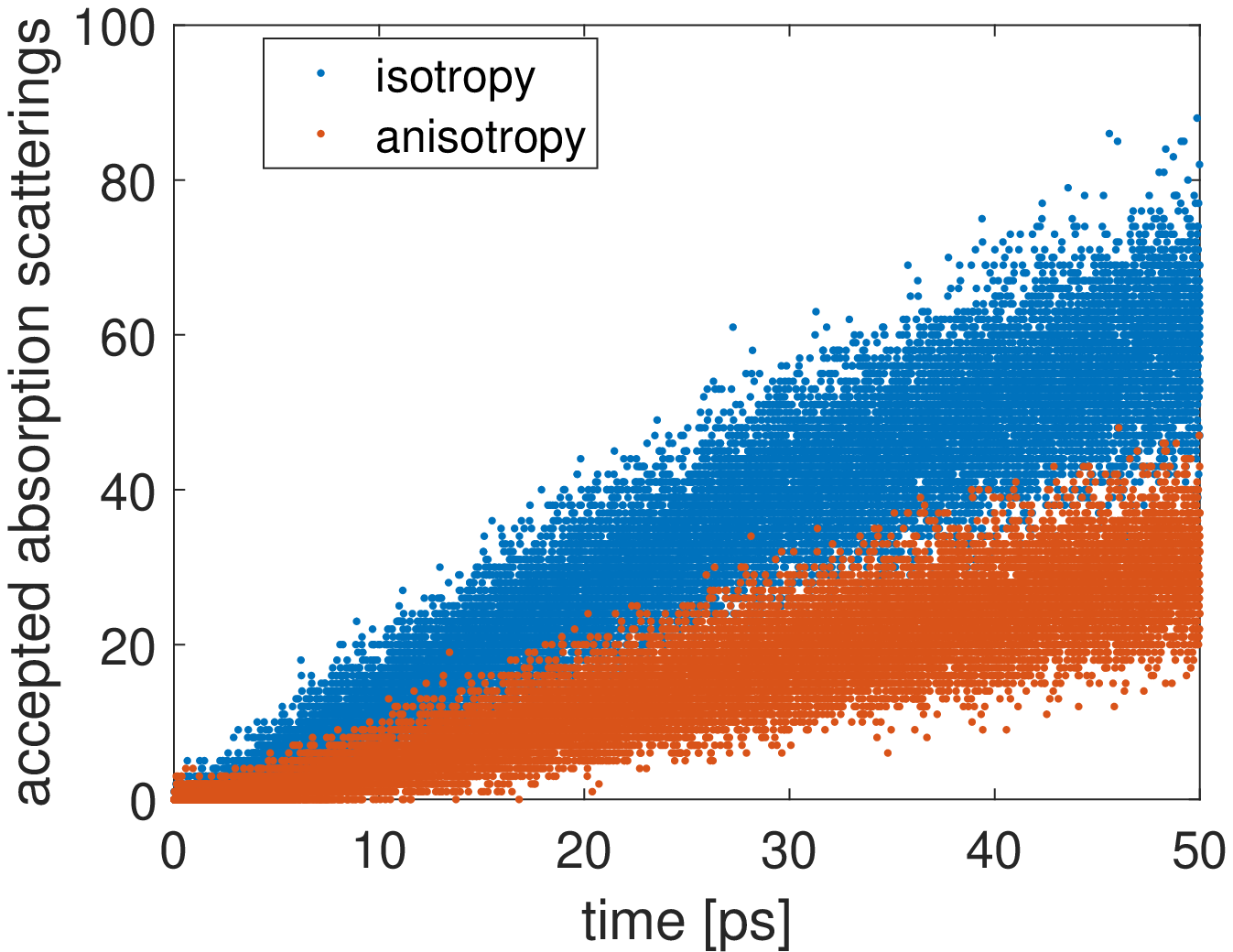}}\\
	\fbox {c)		\includegraphics[width=0.41\columnwidth]{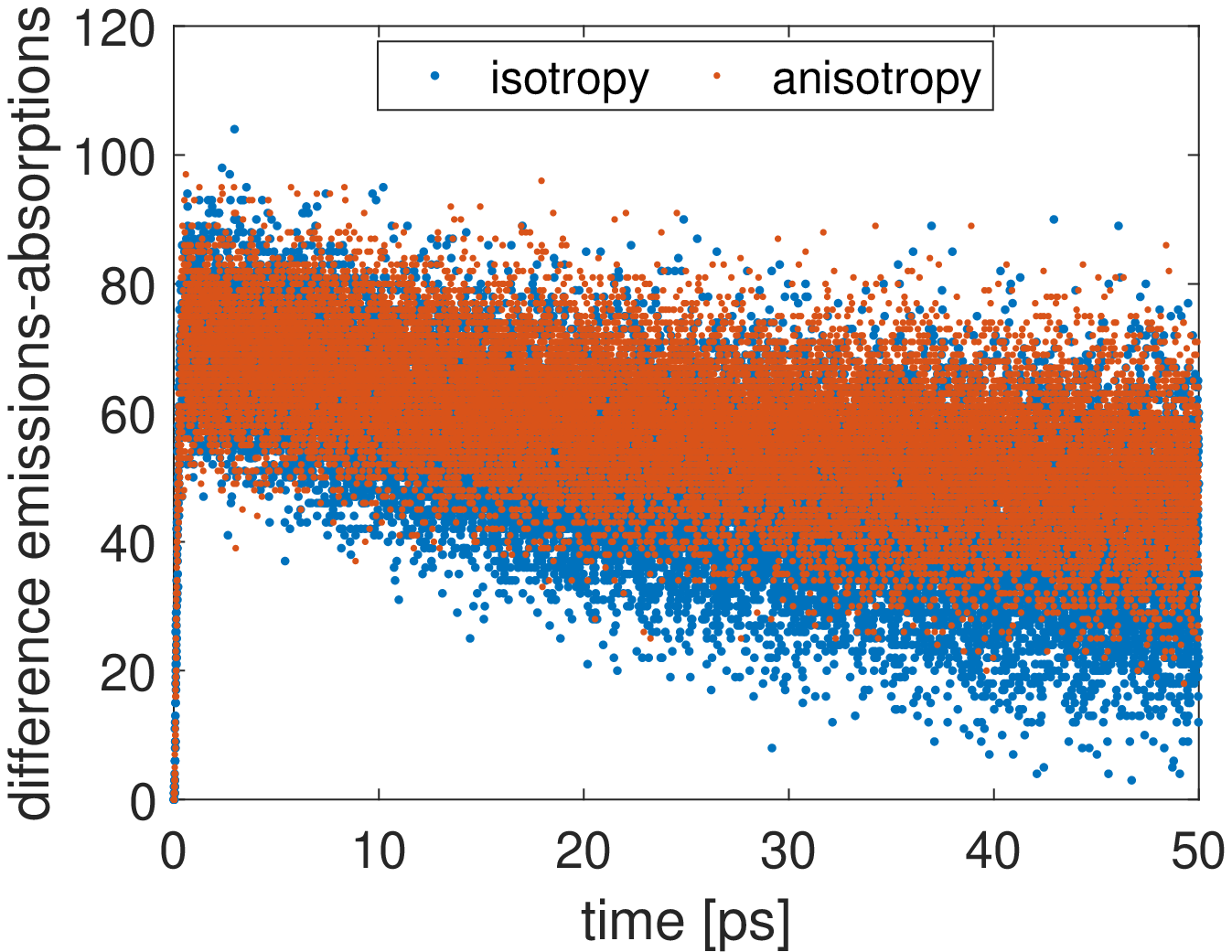}}\\
	\caption{Number of accepted planar optical phonon emission, (a), and absorption, (b), scattering events and their difference, (c), with and without anisotropy. $\varepsilon_F=0.6~{\rm eV}$ and $E=20$ kV/cm.	\label{perc_op}}
\end{figure}

	To explain the effect of the inclusion of the planar phonon anisotropy we will separately analyze the contributions to phonon dynamics due to the two terms of the r.h.s. of Eq.~\eqref{num_sol_ph}, i.e.~the electron--phonon interactions  
\[
C_{\mu-e}=\Delta t \,\,\frac{\alpha_v\beta_{el}}{\left| C_{\nu}\right| } \left(n_{\mu}^{+}({{\mathbf{q}}^{\nu}})-n_{\mu}^{-}({{\mathbf{q}}^{\nu}}) \right),\]
and the phonon--phonon scatterings 
\[
C_{pp}= \Delta t \,\, \frac{g_{\mu}^{(j)} - g_{\mu}^{LE (j)}}{\tau_{\mu}\left(T_{\mu}\left(t^{(j)}\right)\right)},\]
respectively. In Fig. \ref{perc_op}(a) and (b), it is shown that in the isotropic case the number of emitted phonons at each time step is higher than that counted with anisotropy and increases with time, but also the number of absorbed phonons at each time step is higher, and increases with time, without anisotropy. The difference between the number of emitted and absorbed phonons at each time step is reported in Fig. \ref{perc_op}(c); its mean value is lower in the isotropic approximation; this could explain the increasing of electron mean energy in the cases of Fig. \ref{el_T}(a) in which isotropy is more pronounced, because with isotropy the number of electrons which absorb are higher than those which lose quanta of energy. This behavior is present also in the optical $K$ phonons although their total numerical contribution is less relevant. 

The analysis of the number of emitted and absorbed phonons in the electron--phonon part explains the behavior of mean energy of electrons in the isotropic case because during this stage phonons lose a net amount of energy in favor of electrons higher than in the anisotropic situation, but the isotropic lattice local equilibrium temperature should be lower for the same reason, and this is inconsistent with our numerical results; so, it will be fundamental the investigation of phonon--phonon contribution.

\begin{figure}[h!]
	\centering
	\fbox {a)		\includegraphics[width=0.41\columnwidth]{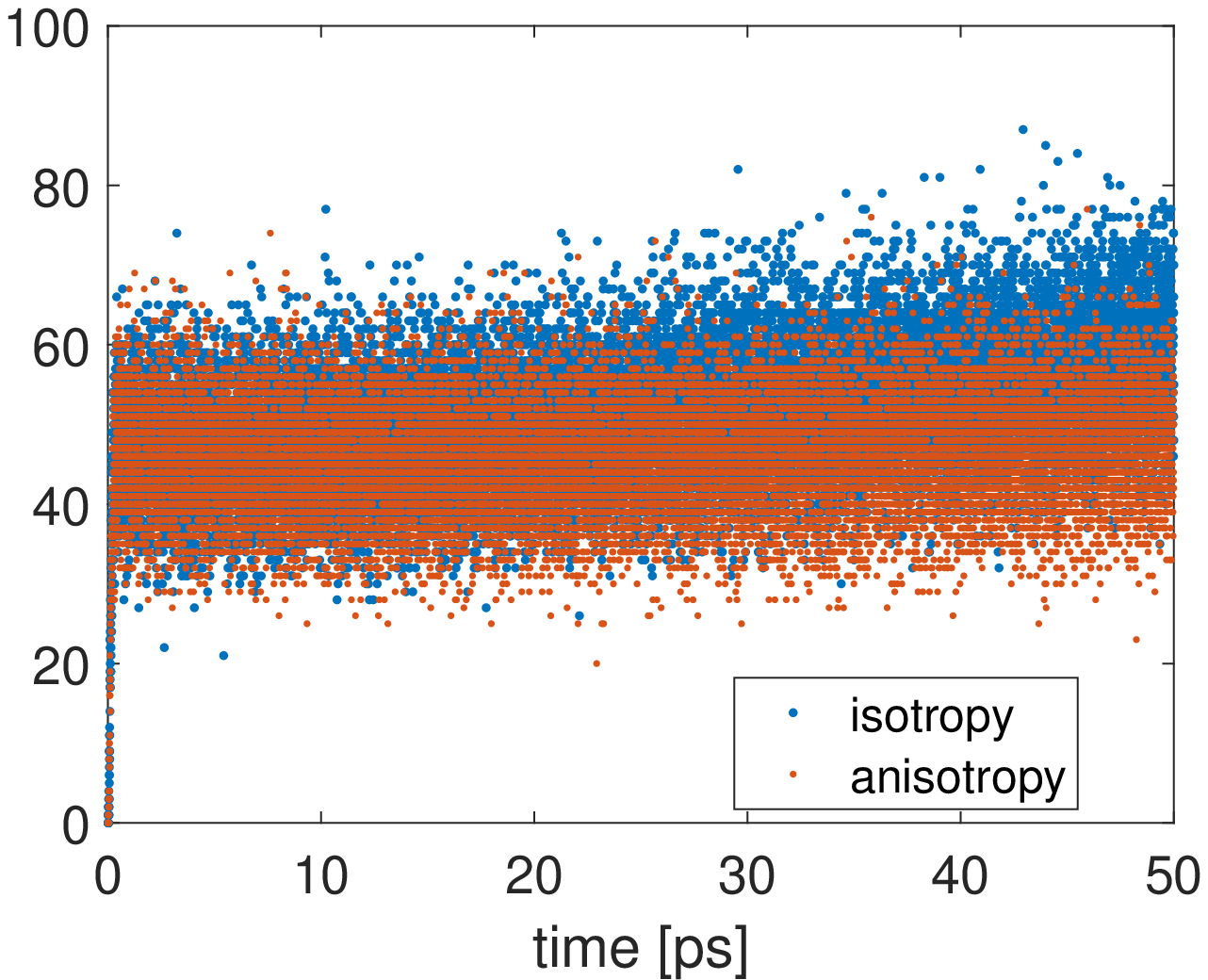}}
	\fbox {b)		\includegraphics[width=0.41\columnwidth]{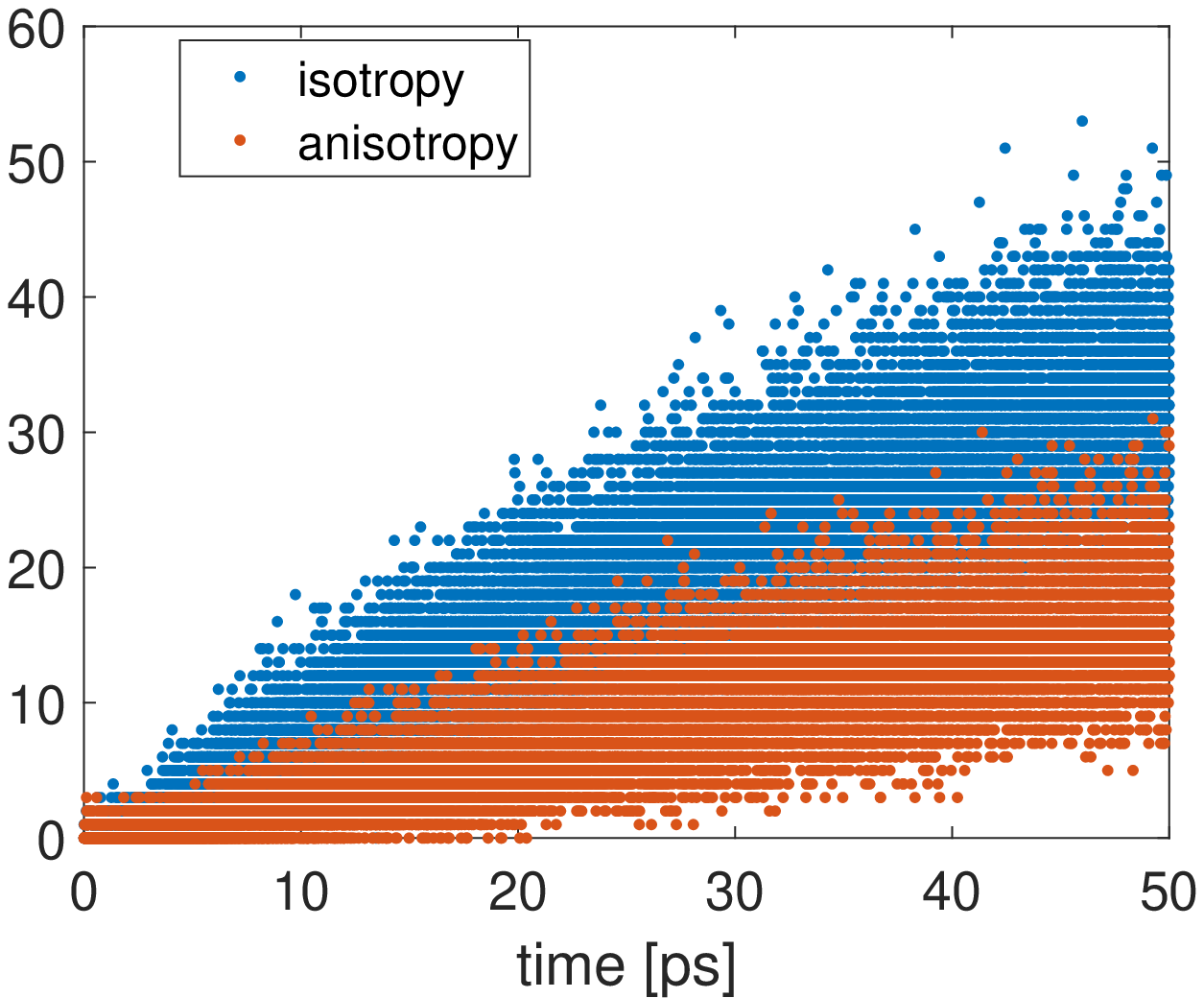}}\\
	\caption{Number of accepted planar optical phonon emission, (a), and absorption, (b), scatterings with negative electron final velocity, with and without anisotropy. $\varepsilon_F=0.6~{\rm eV}$ and $E=20$ kV/cm.	\label{vneg_op}}
\end{figure}

Before, we conclude the analysis of electrical results, in particular the behavior of mean velocity in Fig. \ref{el_T}(b), which is {{the higher the more relevant anisotropy is}}. In Fig. \ref{vneg_op} we show the number of emitted and absorbed phonons in the isotropic and anisotropic situation, respectively, for which the electron final velocity is negative. Also in this subsystem, phonons lose more energy when anisotropy is not included. Both in the emission and absorption part, higher number of electrons acquires negative velocity after a scattering event in the isotropic approximation, and this explains the corresponding lower value of mean electron velocity. $K$ phonons give the same contribution to mean velocity as well.

\begin{figure}[h!]
	\centering
	\fbox {a)		\includegraphics[width=0.41\columnwidth]{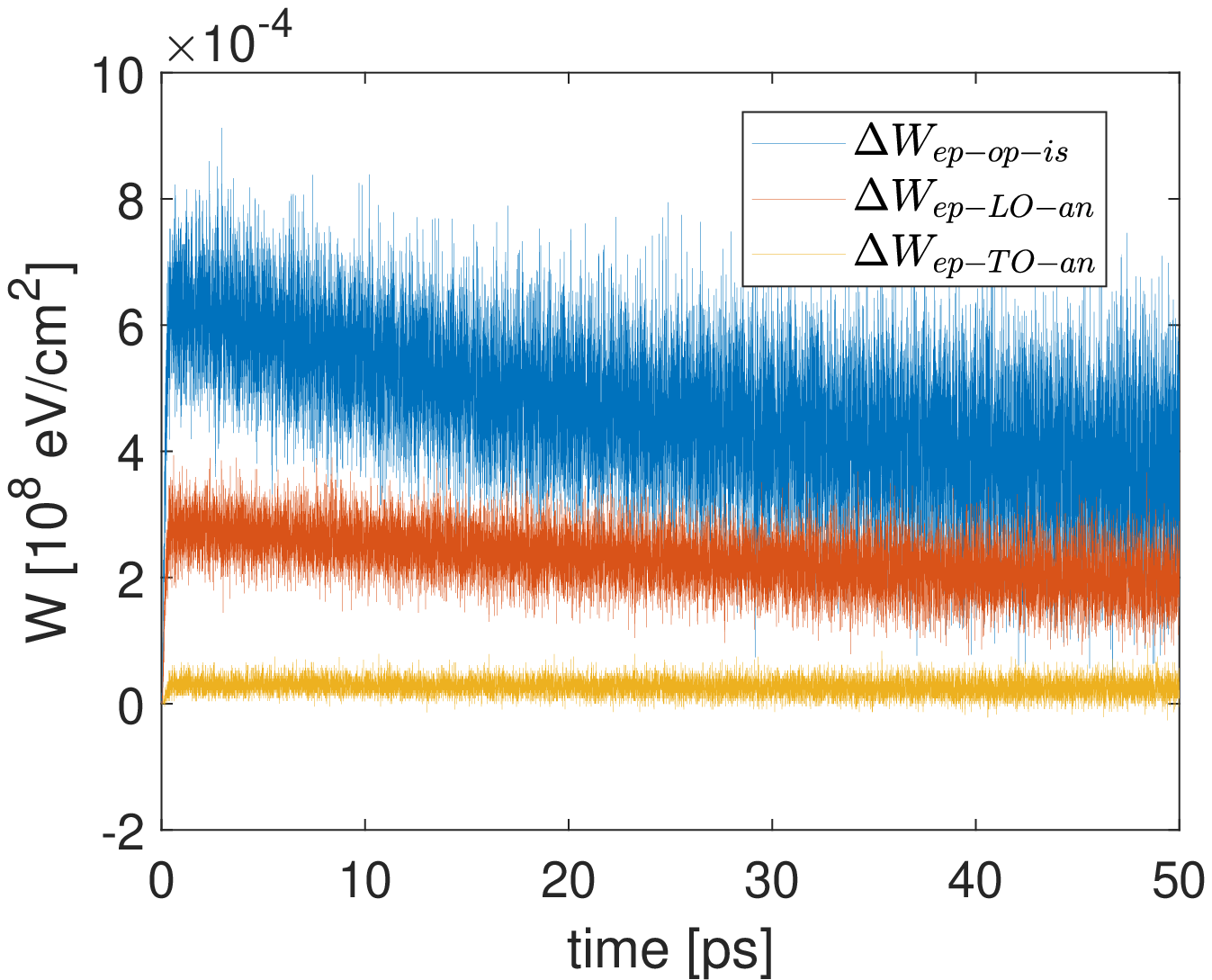}}\\
	\fbox {b)		\includegraphics[width=0.41\columnwidth]{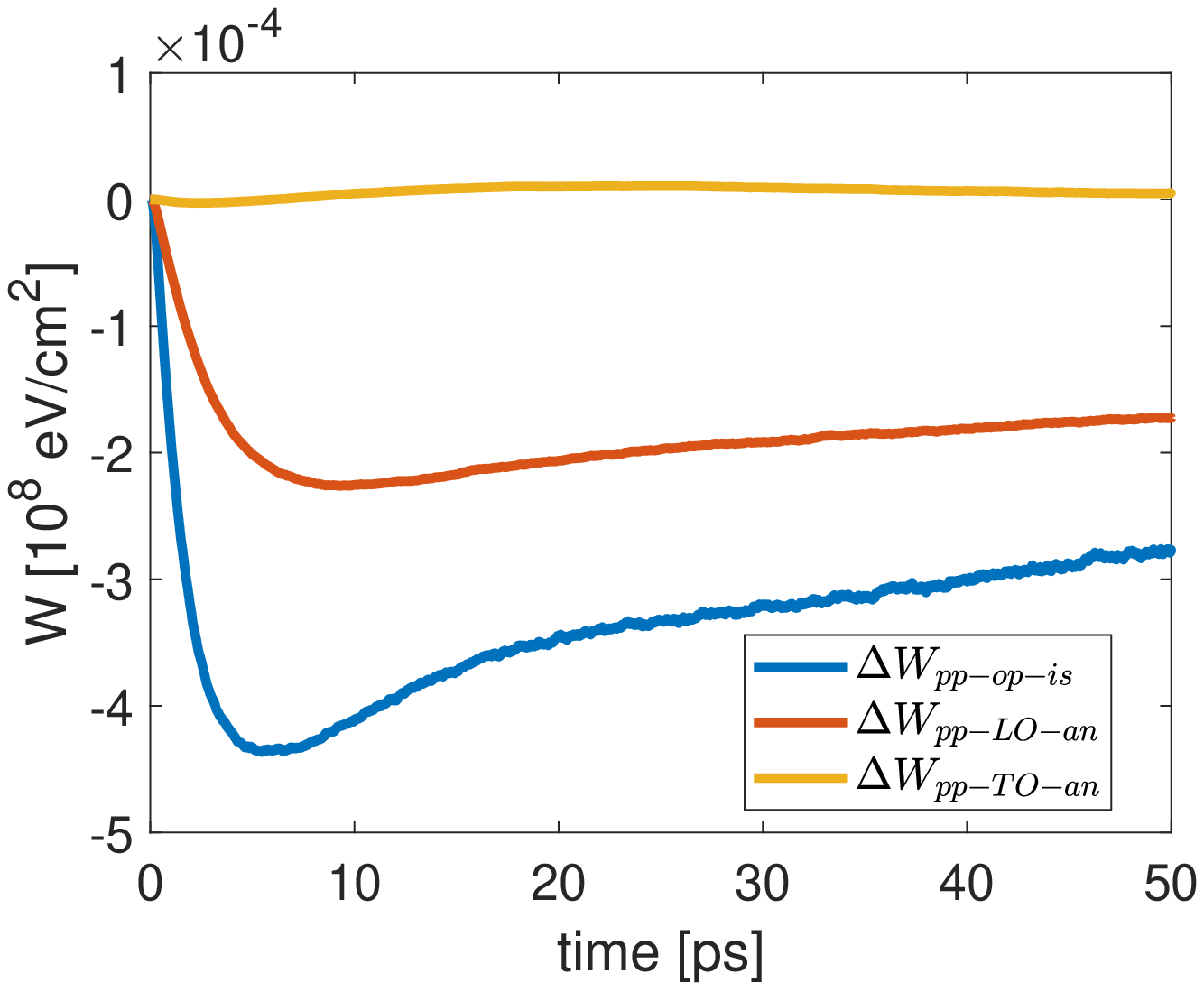}}\\
	\fbox {c)		\includegraphics[width=0.41\columnwidth]{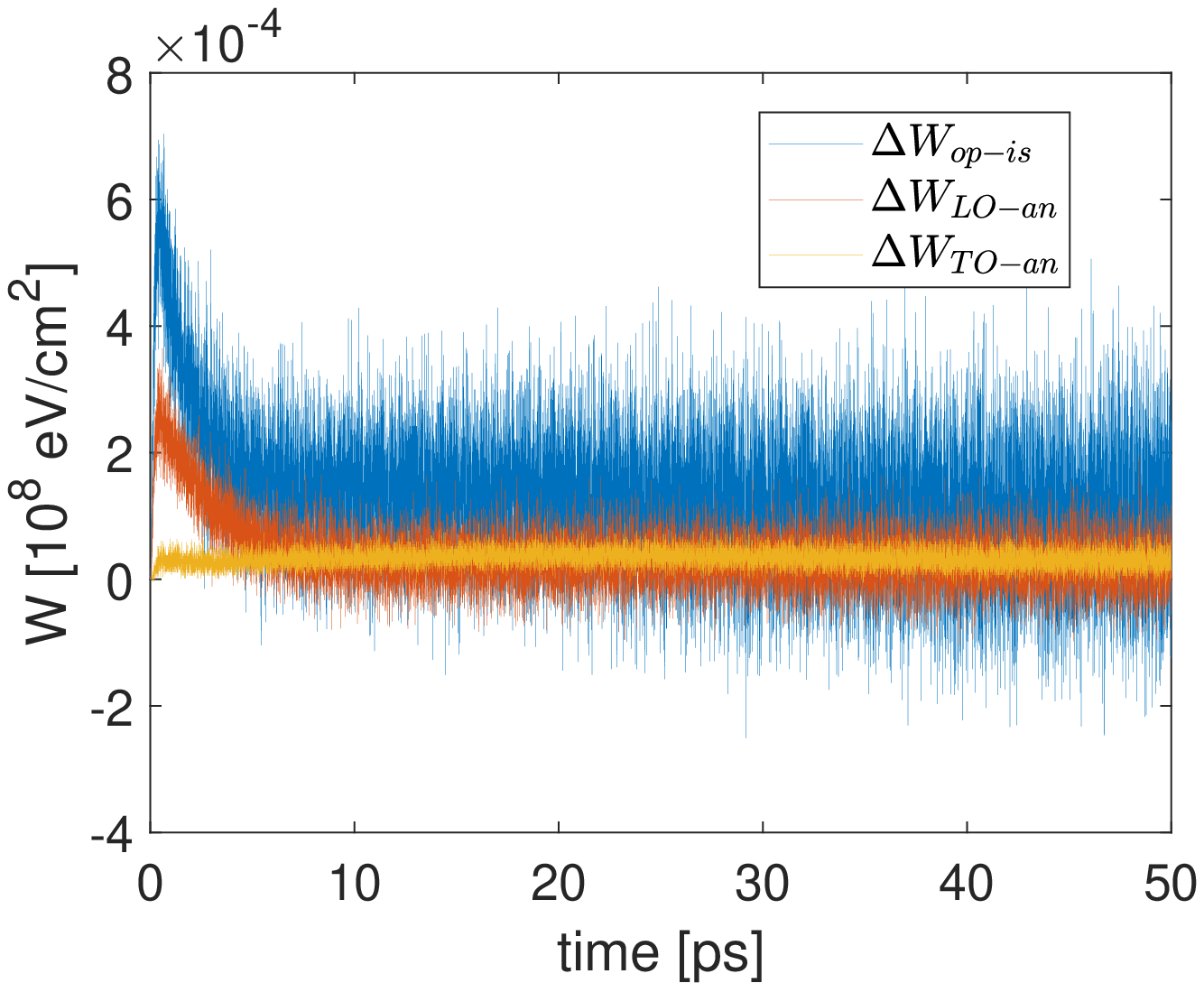}}
	\caption{Planar optical phonon energy density variation at each time step for the electron--phonon part (a), phonon--phonon part (b), total (c), with (an) and without (is) anisotropy. $\varepsilon_F=0.6~{\rm eV}$ and $E=20$ kV/cm.	\label{W_ph1_pp}}
\end{figure}

\begin{figure}[h!]
	\centering
	\includegraphics[width=0.6\columnwidth]{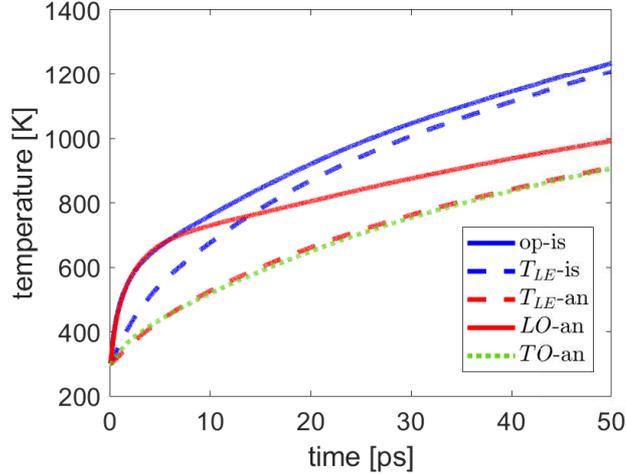}\\
	\caption{Local equilibrium temperature $T_{LE}$ and optical phonons temperatures with (an) and without (is) anisotropy, with the first definition of $T_{LE}$. $\varepsilon_F=0.6~{\rm eV}$ and $E=20$ kV/cm.	\label{T_comp2}}
\end{figure}

\begin{figure}[h!]
	\centering
	\fbox {a)		\includegraphics[width=0.41\columnwidth]{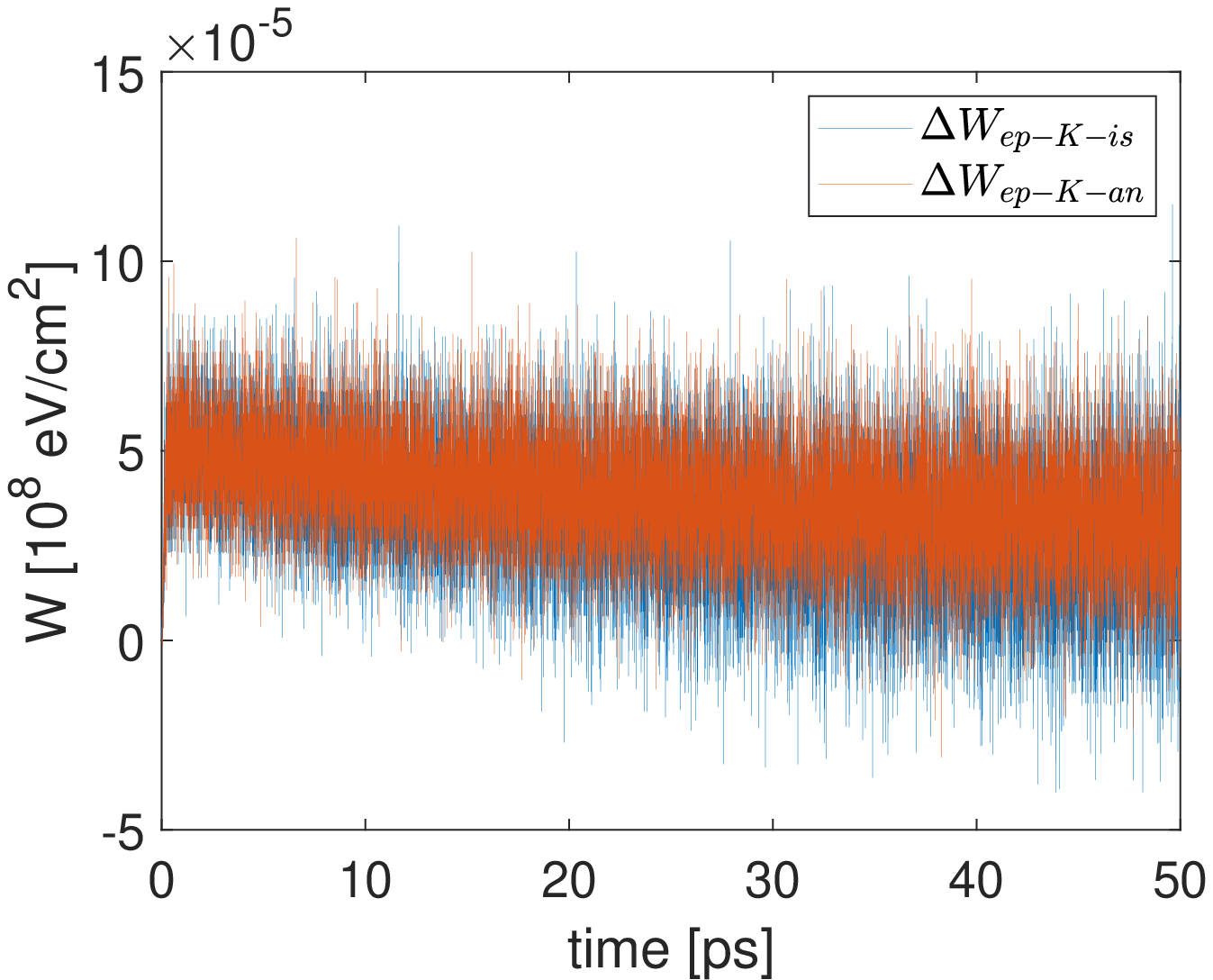}}\\
	\fbox {b)		\includegraphics[width=0.41\columnwidth]{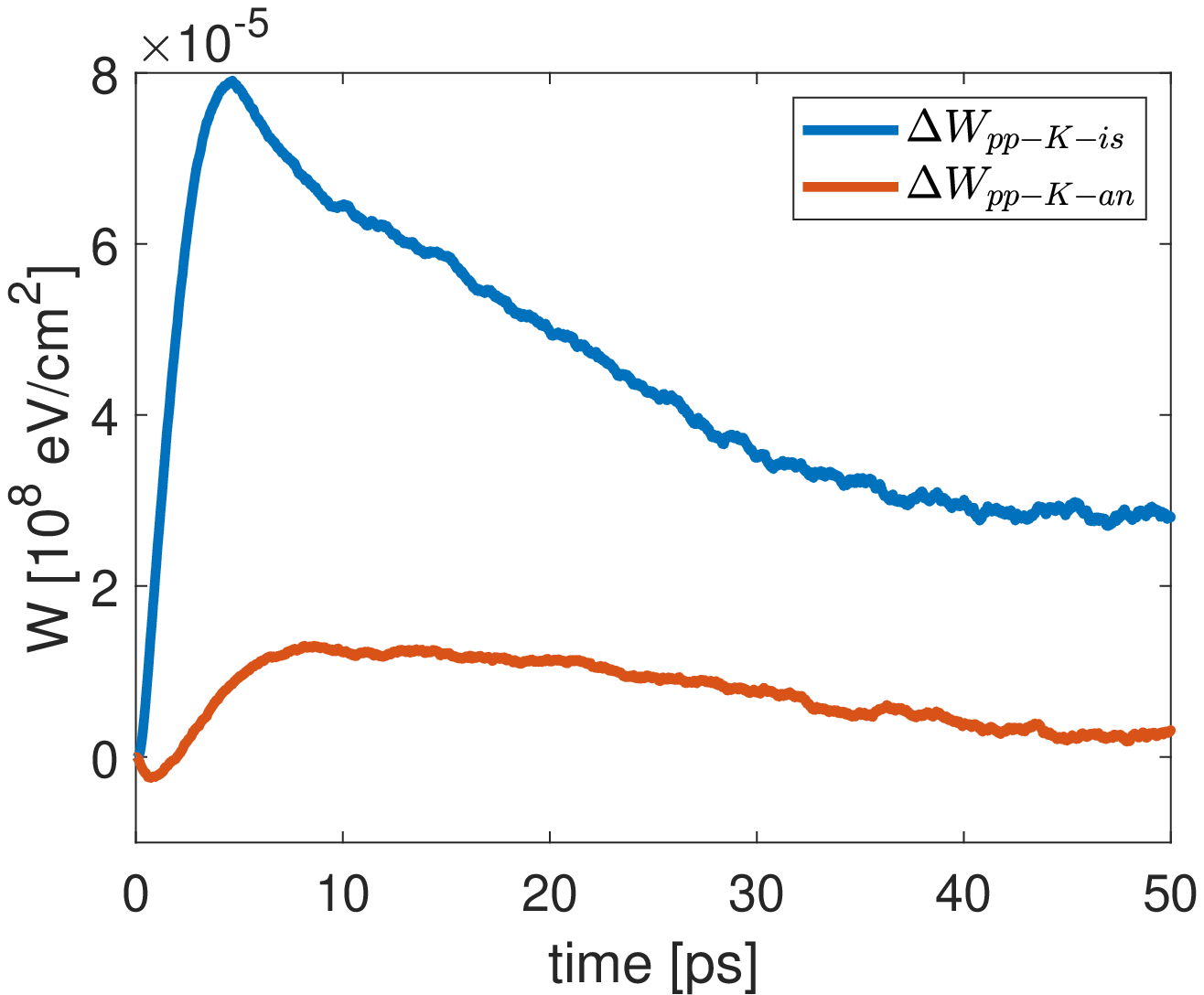}}\\
	\fbox {c)		\includegraphics[width=0.41\columnwidth]{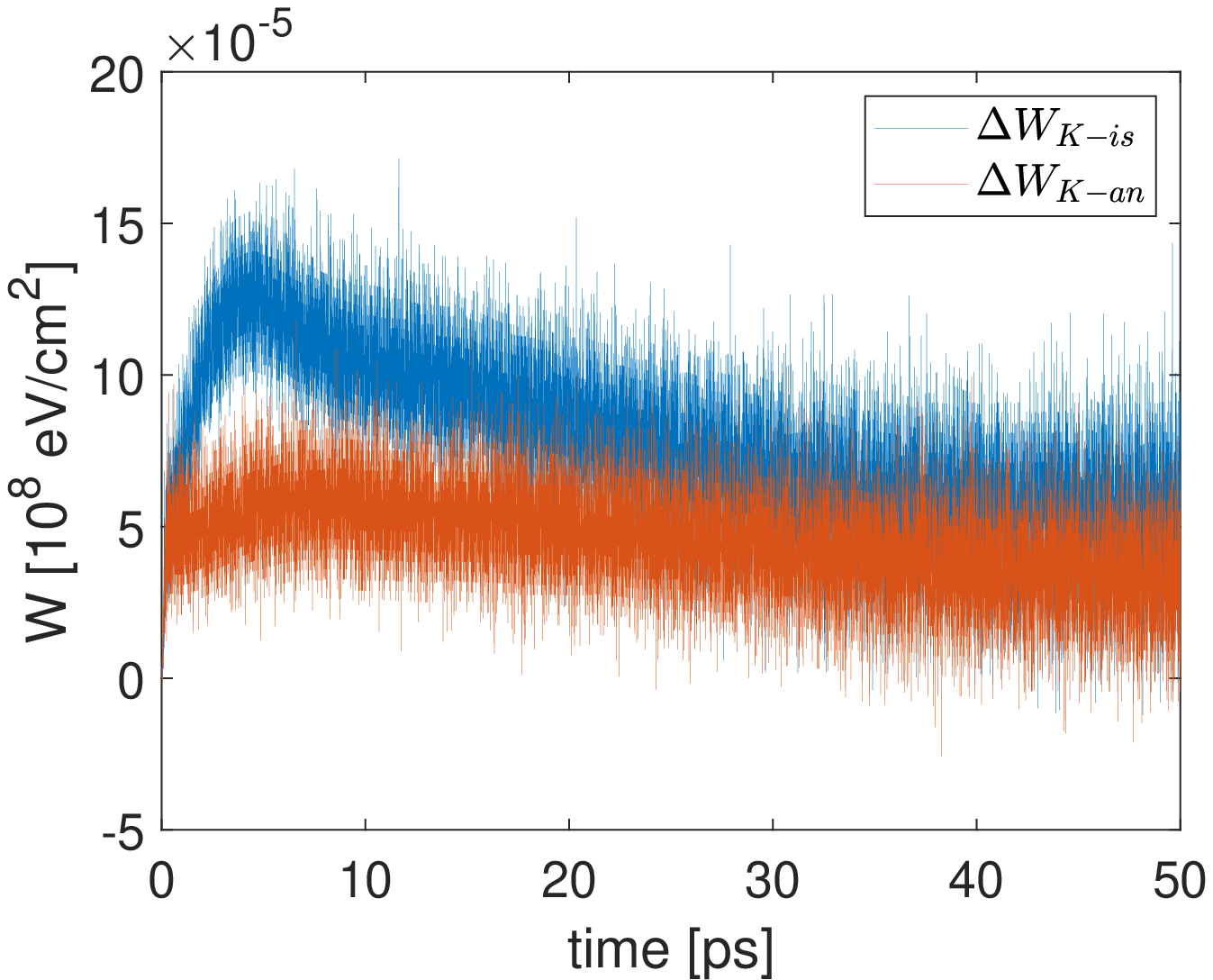}}
	\caption{$K$ phonon energy density variation at each time step for the electron--phonon part (a), phonon--phonon (b), total (c), with (an) and without (is) anisotropy. $\varepsilon_F=0.6~{\rm eV}$ and $E=20$ kV/cm.	\label{W_ph2_pp}}
\end{figure}

{{Regarding the phonon--phonon part,}} at each time step, we compare the contribution to total energy density given by electron--phonon and phonon--phonon scatterings, $\Delta W_{ep}$ and $\Delta W_{pp}$, respectively; {{they are computed by inserting terms $C_{\mu-e}$ and $C_{pp}$ into the Eq.~\eqref{num_mu}.}} In Fig. \ref{W_ph1_pp}, we consider the results about optical phonons which are the most effective; in the isotropic case, panel (a), the energy density due to the electron--phonon step reaches higher values than those of phonon--phonon interactions, panel (b), which give lower partial contribution; their difference, panel (c), which is related to the r.h.s of Eq.~\eqref{ph_tr}, is much higher with isotropy, explaining the fact that the temperature is higher when planar phonon anisotropy is not included. The energy density due to phonon--phonon part is negative for planar optical isotropic phonons and $LO$ anisotropic phonons because their temperature is higher than the local equilibrium one as shown in Fig. \ref{T_comp2}. Instead, for $K$ phonons, $\Delta W_{ep-K}$ has about the same value with and without anisotropy while $\Delta W_{pp-K}$ is higher in the isotropic case, as reported in Fig. \ref{W_ph2_pp}(a) and (b). Also for $K$ phonons, the net variation $\Delta W_{K}$ is higher for isotropy (Fig. \ref{W_ph2_pp}c), consistently with the higher values of temperatures and energy densities in this case. 

\begin{figure}[h!]
	\centering
	\fbox{a)\includegraphics[width=0.41\columnwidth]{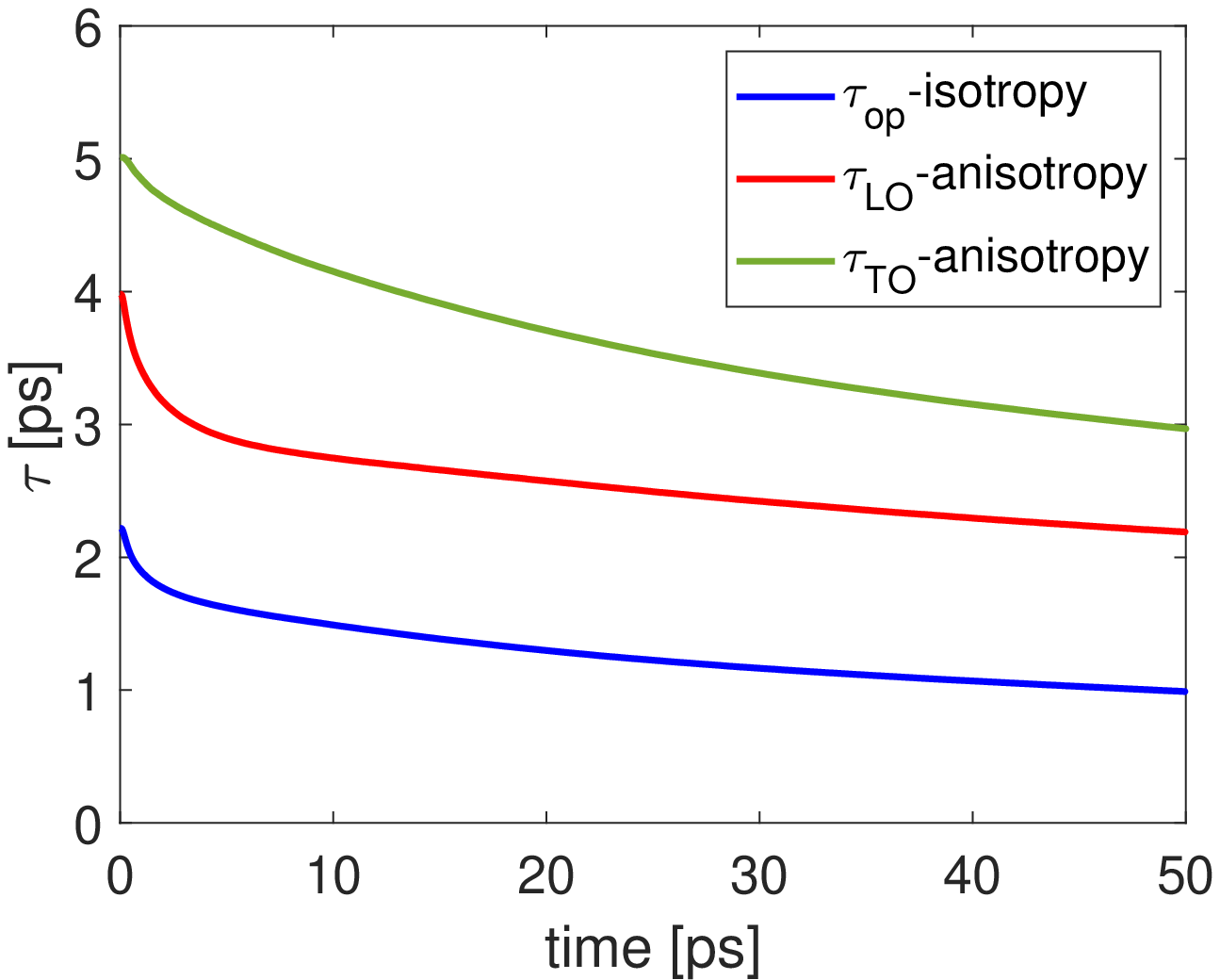}}
	\fbox{b)\includegraphics[width=0.41\columnwidth]{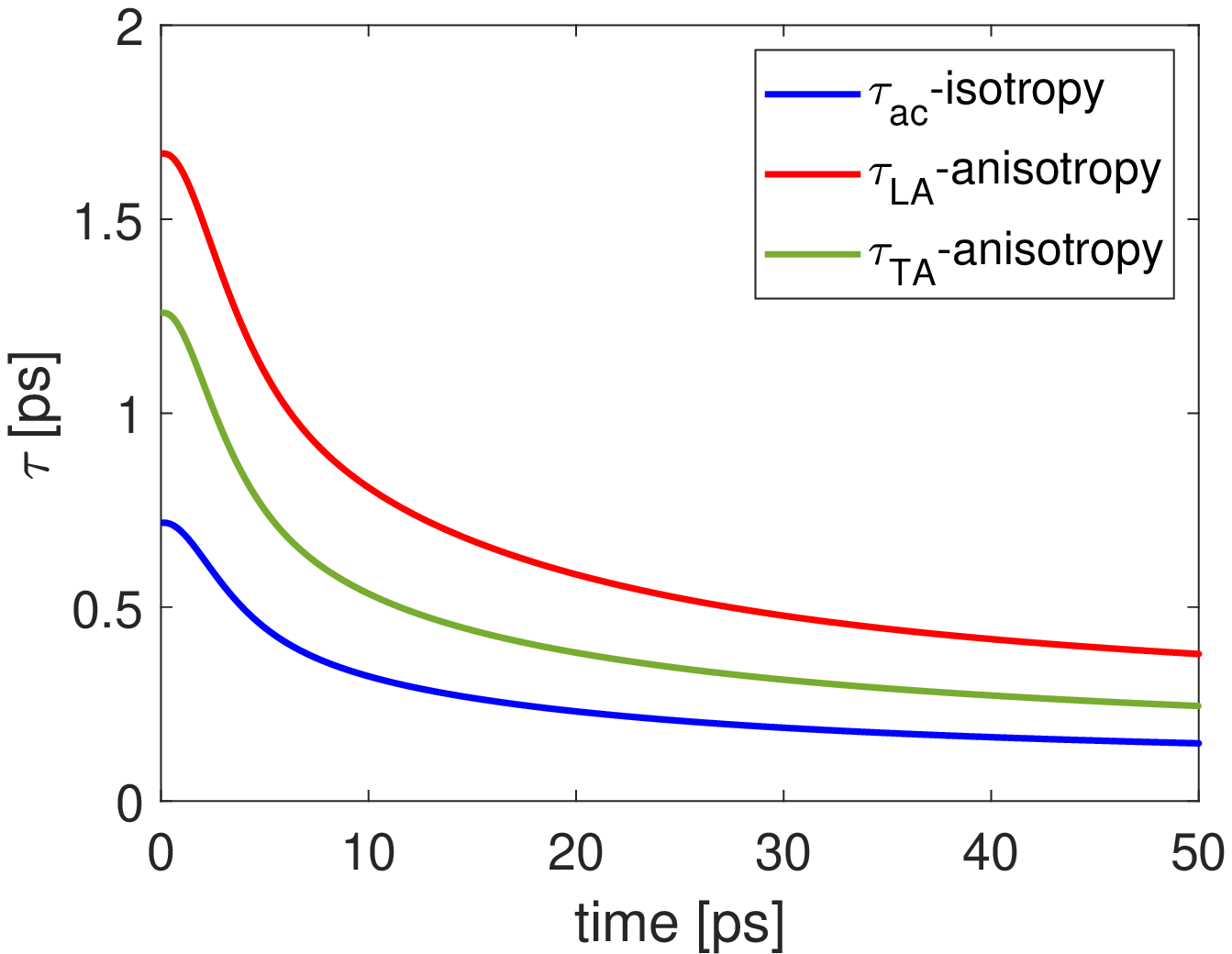}}\\
	\caption{Relaxation time of the whole planar optical branch $(LO+TO)$ in the isotropic case, and $LO$, $TO$ relaxation times in the anisotropic case (a), and of $(LA+TA)$, $LA$, $TA$ phonons, (b). $\varepsilon_F=0.6~{\rm eV}$ and $E=20$ kV/cm.	\label{tau_op}}
\end{figure}

The behavior of $\Delta W_{ep-op}$ in Fig. \ref{W_ph1_pp}(a) is counter-intuitive considering only the previous results on the net difference between the number of emitted and absorbed phonons in the isotropic case, {{for which there is a slight prevalence of absorption processes with respect to the anisotropic situation and then the corresponding phonon energy density is expected to be lower than the anisotropic one}}; this result should be explained as an evident consequence of the isotropic approximation because, even if their number is lower, the emitted phonons reach more energetic regions than the anisotropy would allow. We remind that in the isotropic approximation the angular dependence of the optical scattering matrix elements disappears at all. Moreover, we have different temperature dependent relaxation times which enter Eq.~\eqref{prod_ph} for the evaluation of $T_{LE}$ and which in the isotropic case are considered just by means of the Matthiessen rule giving at $50$ ps an optical relaxation time for the unique $(LO+TO)$ population about half and a third of the separated $LO$ and $TO$ phonons, respectively, introducing an overestimation of heating effects (Fig. \ref{tau_op}a). Planar acoustic phonons relaxation times have a similar behavior (Fig. \ref{tau_op}b).

The previous results are obtained by using the first definition of the lattice local equilibrium temperature $T_{LE}$ given in Eq.~\eqref{prod_ph}. In the following, we will repeat our investigation by means of the second definition of $T_{LE}$ given in Eq.~\eqref{second_T}, in which we remind that the anisotropy related to the relaxation times do not explicitly enter. 

\begin{figure}[h!]
	\centering
	\fbox {a)	\includegraphics[width=0.41\columnwidth]{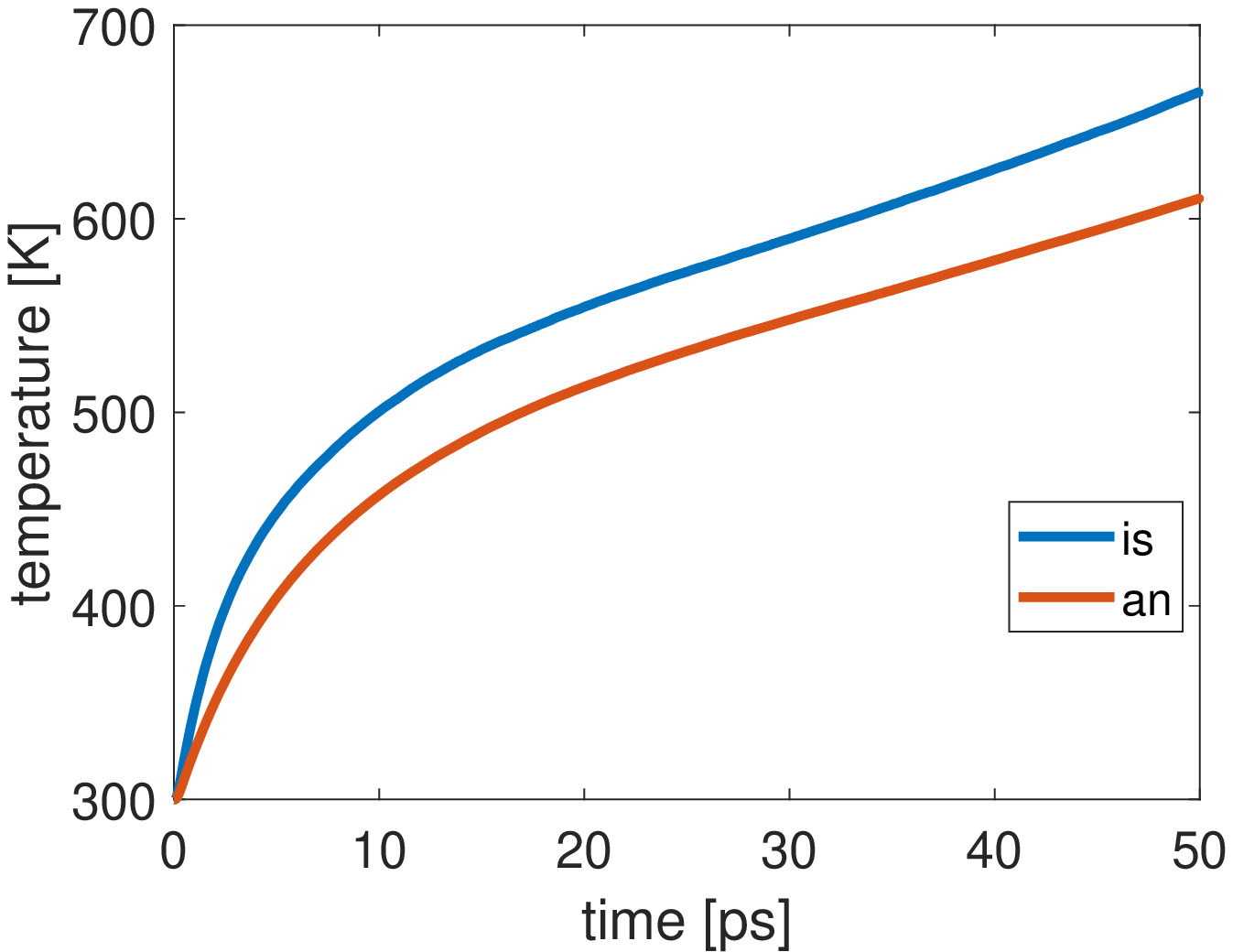}}
	\fbox {b)	\includegraphics[width=0.41\columnwidth]{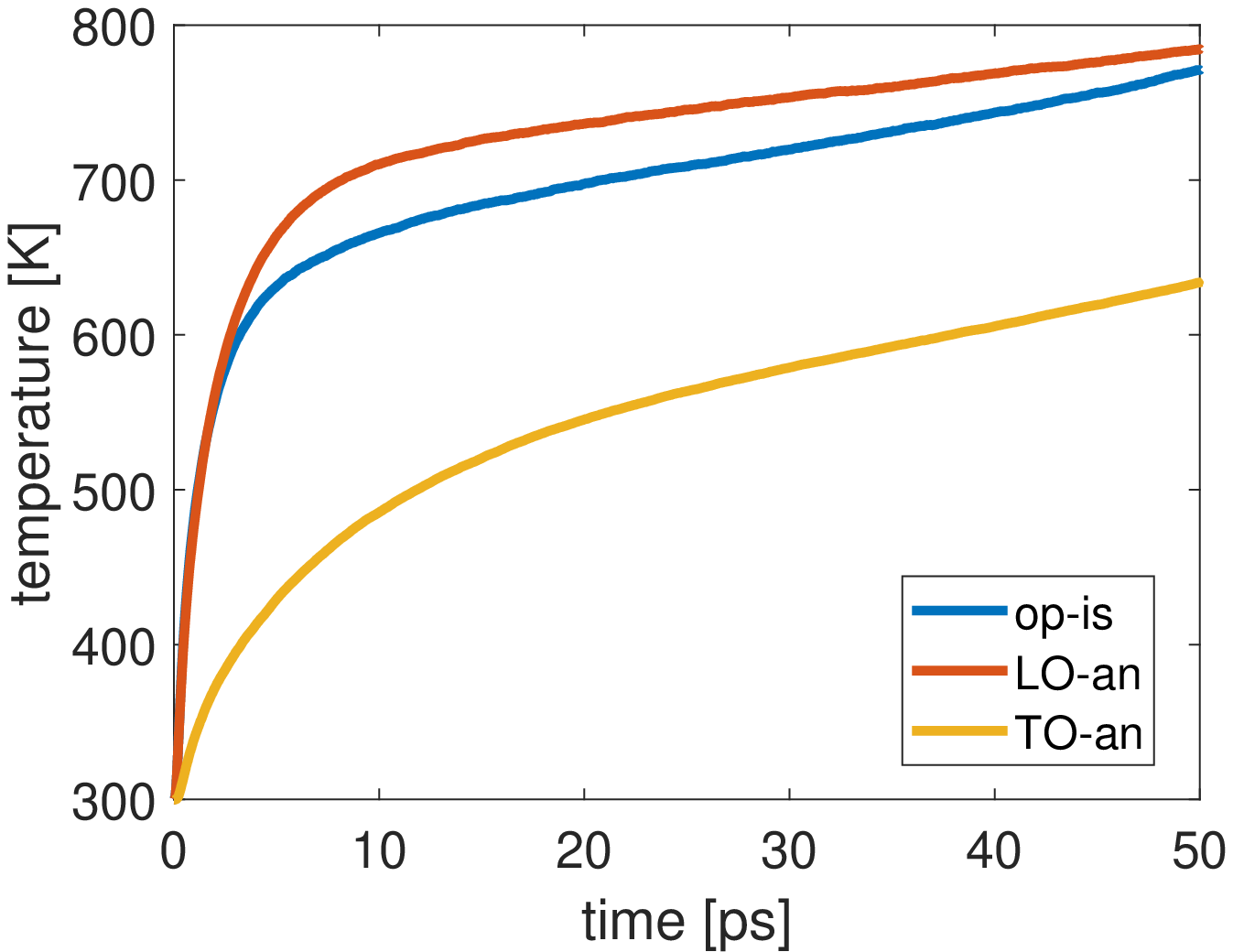}}\\
	\caption{Local equilibrium temperature, (a), optical phonons temperatures, (b), with (an) and without (is) anisotropy, when the second definition of $T_{LE}$ is used. $\varepsilon_F=0.6~{\rm eV}$ and $E=20$ kV/cm.	\label{T_comp3}}
\end{figure}

As shown in Fig. \ref{T_comp3}(a), the local equilibrium temperature $T_{LE}$ is higher with isotropic approximation also when its second definition in Eq.~\eqref{second_T} is used; {{the discrepancy between isotropic and anisotropic cases is lower with respect to that given by the first definition in Eq.~\eqref{prod_ph} and it is of}} about 8.5\% ad not of about 33\%; moreover, the isotropic curve presents a more evident change in the curvature at long times. In Fig. \ref{T_comp3}(b), temperatures of planar optical phonons have a different behavior {with the second definition of $T_{LE}$}; $TO$ phonons have again the lowest value, but the anisotropic $LO$ temperature is higher than the isotropic $(LO+TO)$ one and they tend to thermalize to a common value at 50 ps (see Fig. \ref{T_comp2} for comparison).

\begin{figure}[h!]
	\centering
	\fbox {a)		\includegraphics[width=0.41\columnwidth]{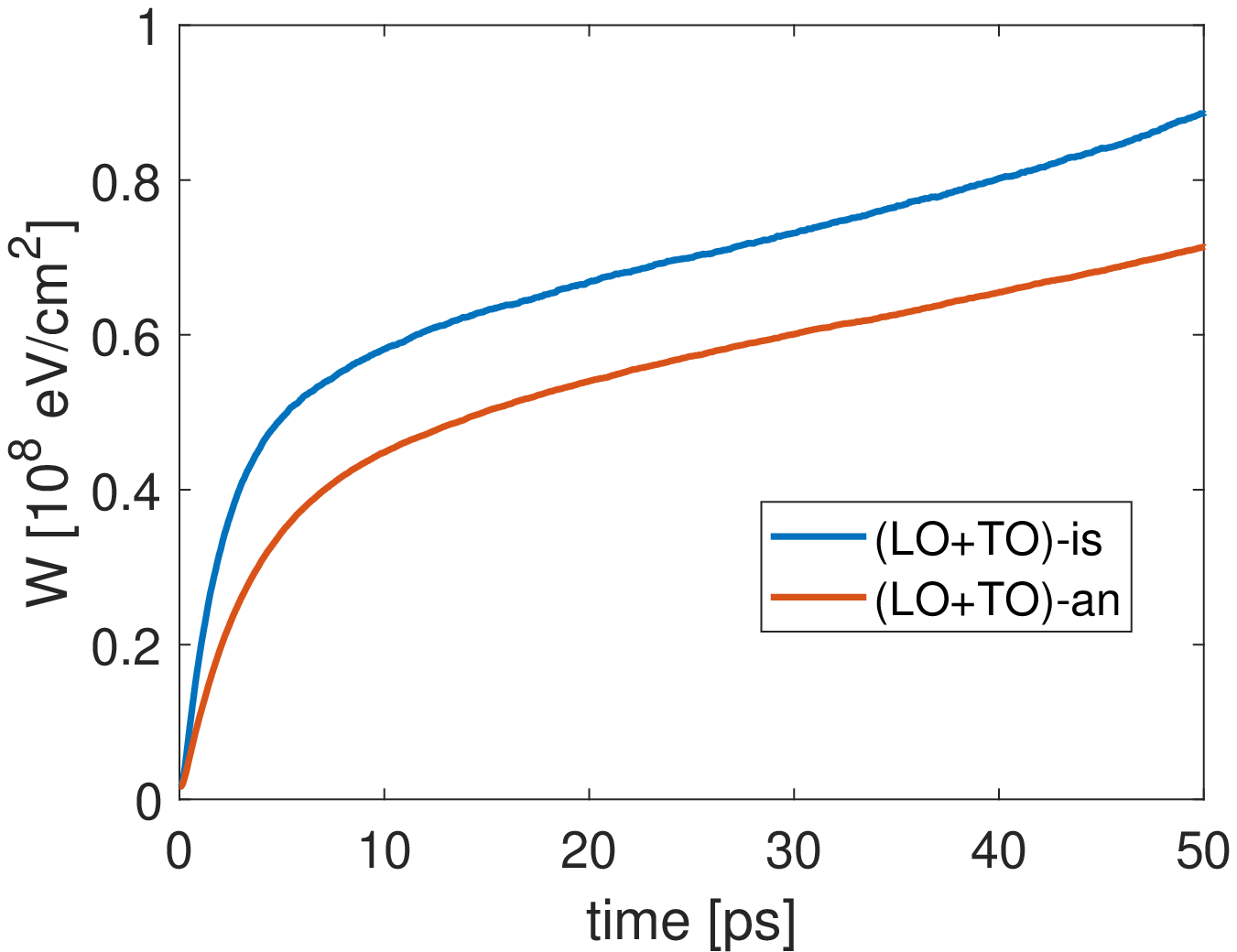}}
	\fbox {b)		\includegraphics[width=0.41\columnwidth]{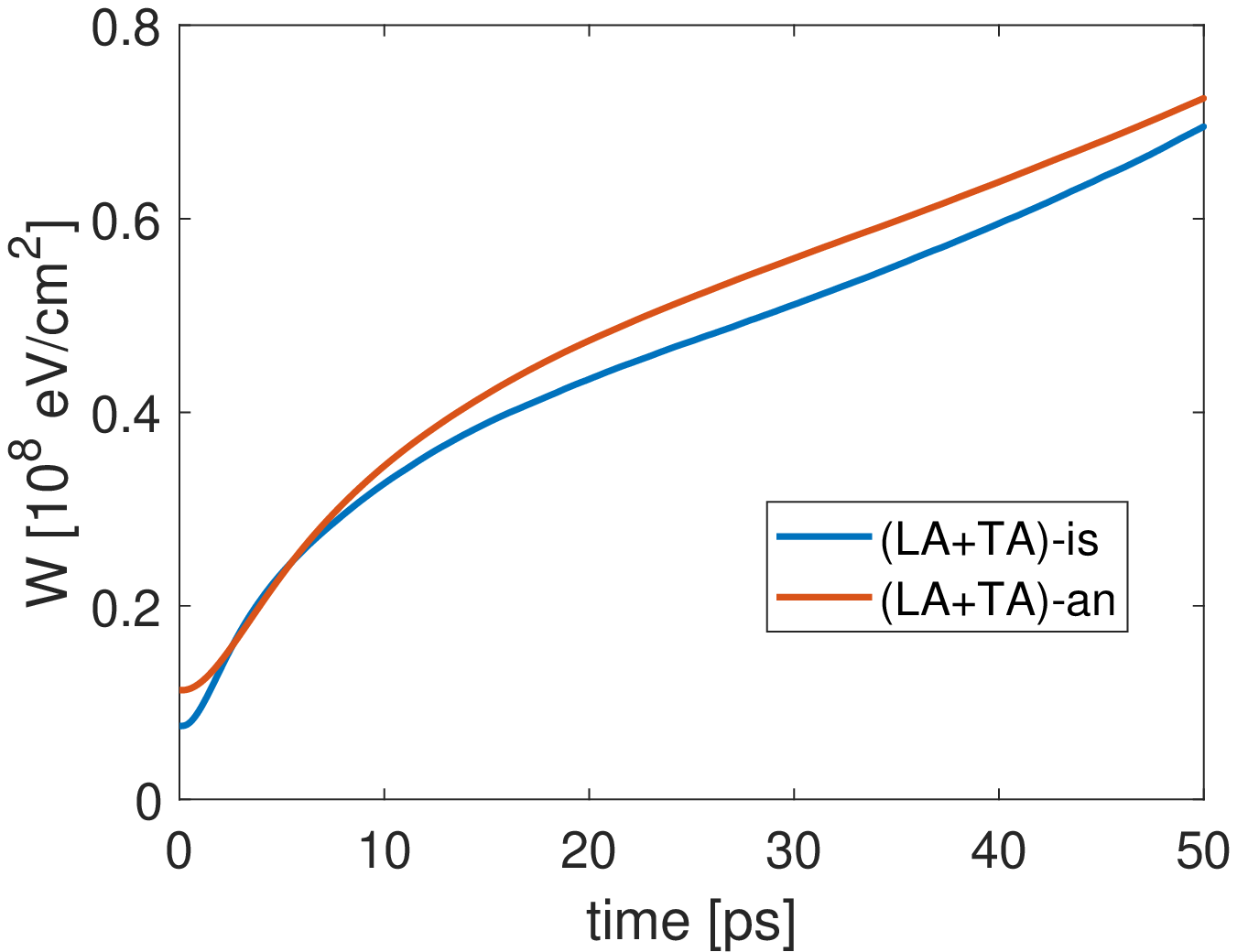}}\\
	\caption{Planar phonon energy density, with (an) and without (is) anisotropy, when the second definition of $T_{LE}$ is used. $\varepsilon_F=0.6~{\rm eV}$ and $E=20$ kV/cm.	\label{T_p2}}
\end{figure}

These results are coherent with time evolution of planar optical phonon energy density (Fig. \ref{T_p2}); interestingly, with the second definition of $T_{LE}$, acoustic $(LA+TA)$ phonon energy densities have a different behavior in the curvature after 10 ps but with comparable numerical values which seem to become almost equal at 50 ps.

\begin{figure}[h!]
	\centering
	\includegraphics[width=0.6\columnwidth]{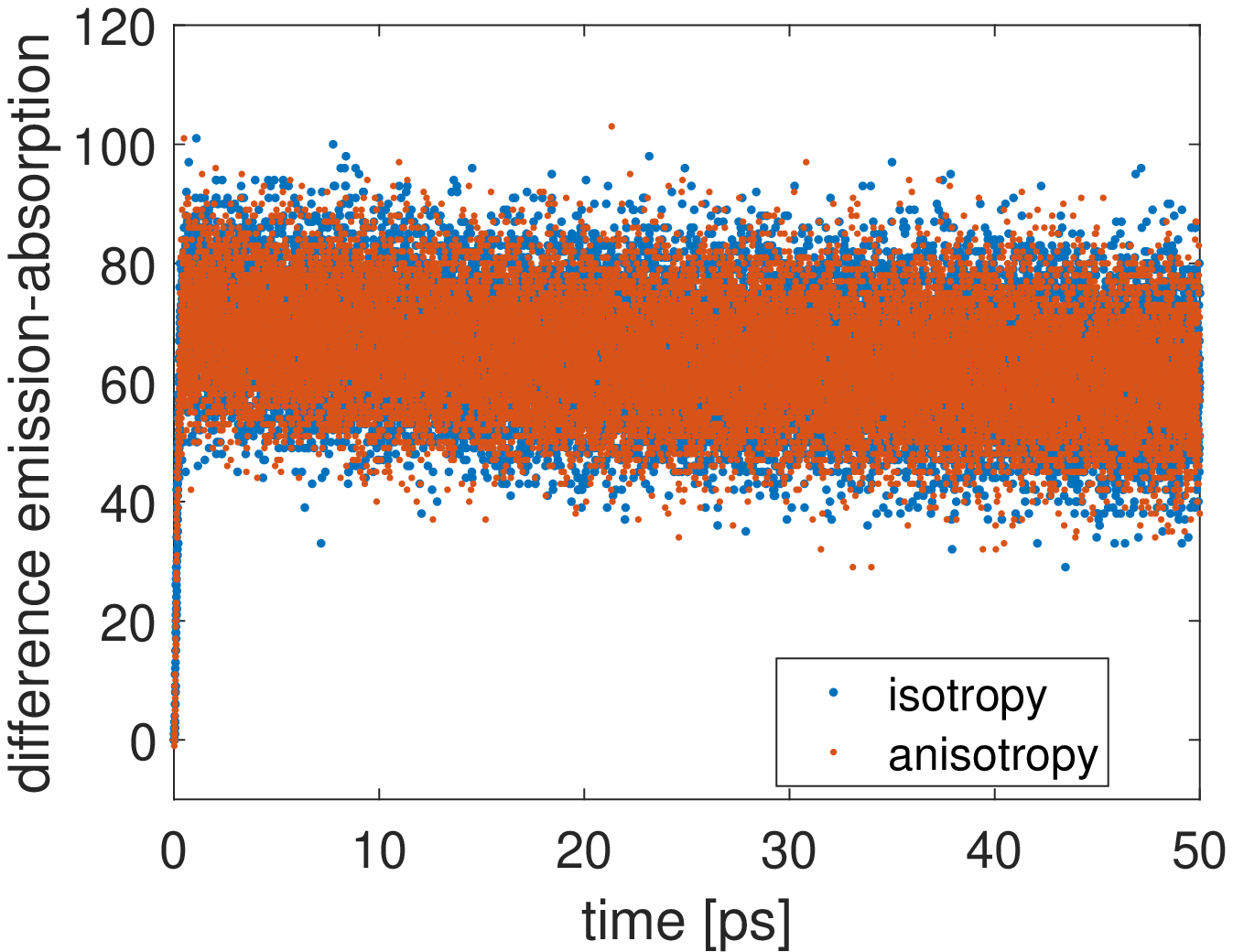}
	\caption{Difference between number of accepted planar optical phonon emissions and absorptions with and without anisotropy, when the second definition of $T_{LE}$ is used. $\varepsilon_F=0.6~{\rm eV}$ and $E=20$ kV/cm.	\label{perc_op3}}
\end{figure}

\begin{figure}[h!]
	\centering
	\fbox {a)		\includegraphics[width=0.41\columnwidth]{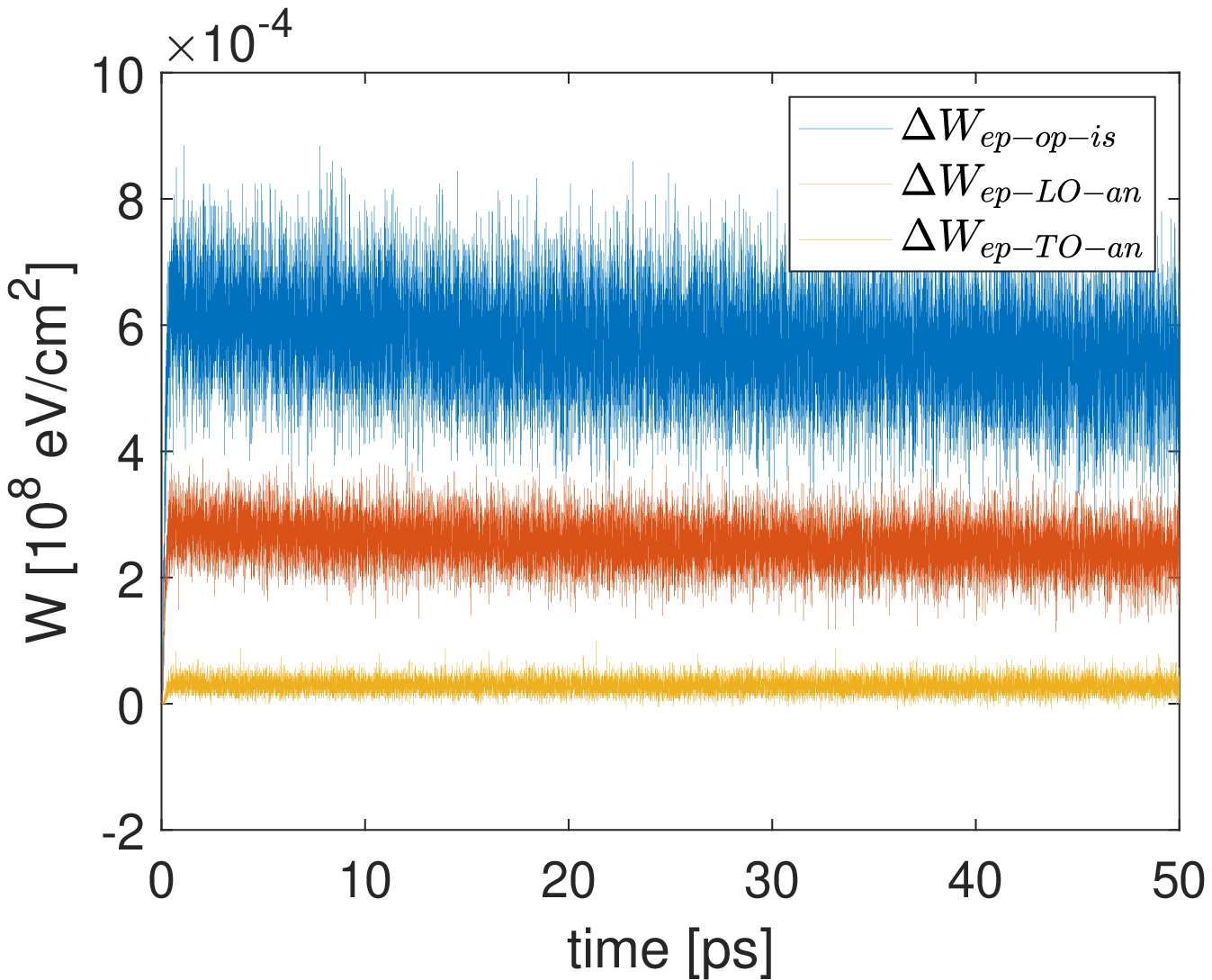}}\\
	\fbox {b)		\includegraphics[width=0.41\columnwidth]{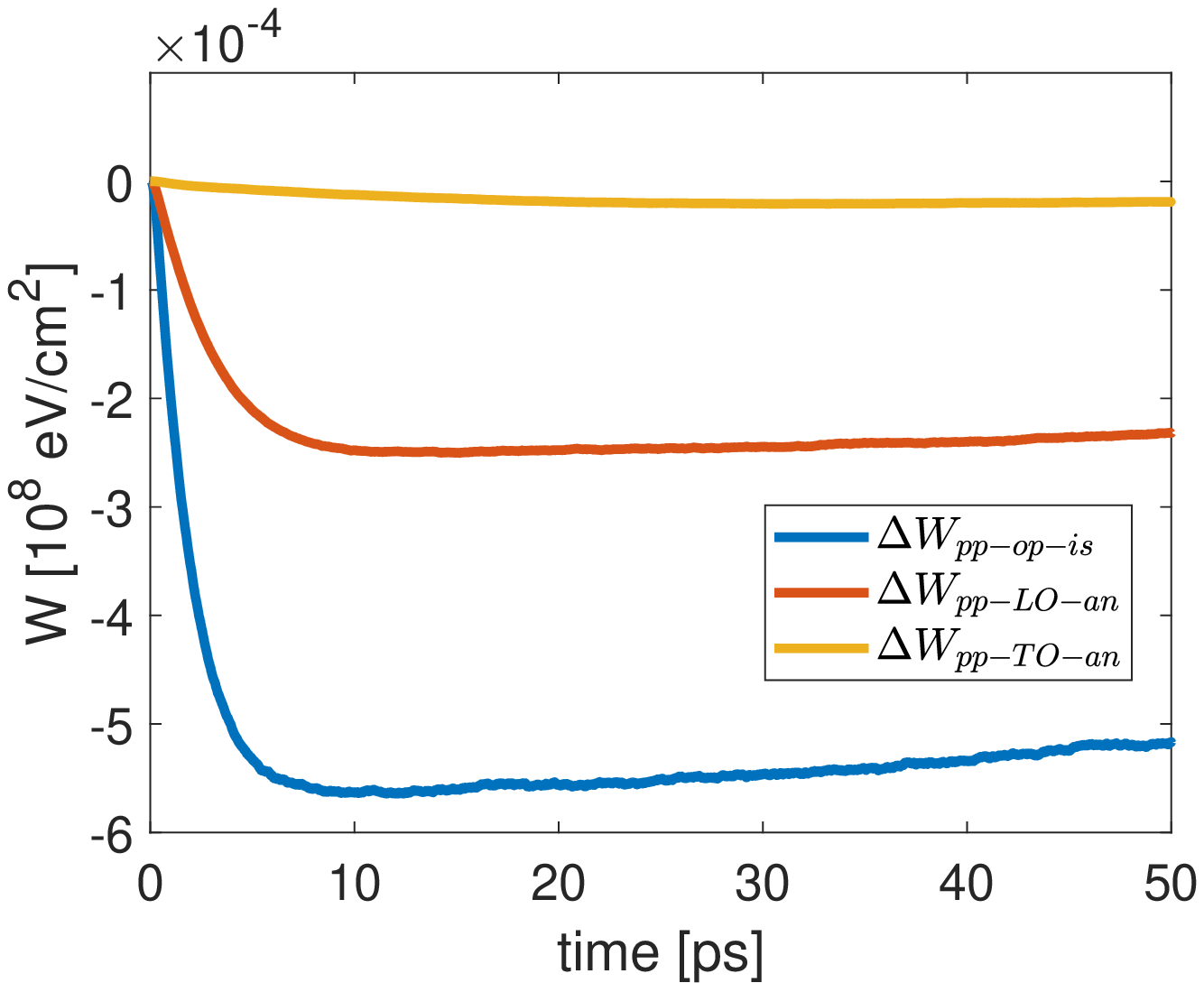}}\\
	\fbox {c)		\includegraphics[width=0.41\columnwidth]{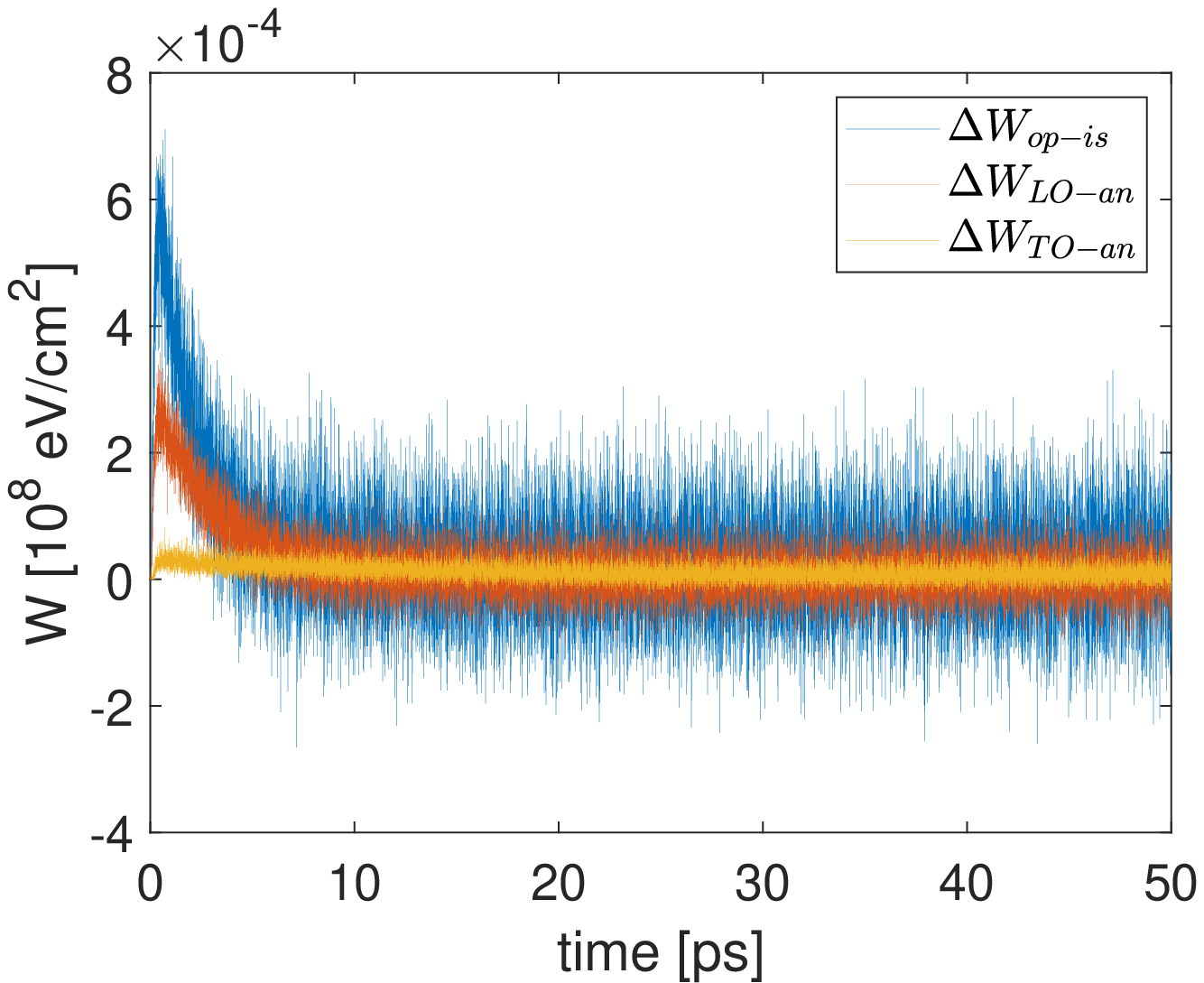}}
	\caption{Planar optical phonon energy density variation at each time step for the electron--phonon part (a), phonon--phonon part (b), total (c), with (an) and without (is) anisotropy, when the second definition of $T_{LE}$ is used. $\varepsilon_F=0.6~{\rm eV}$ and $E=20$ kV/cm.	\label{W_ph3_pp}}
\end{figure}

With the second definition of $T_{LE}$, the difference between the number of emission and absorption processes involving optical phonons is equal for isotropic and anisotropic case (Fig. \ref{perc_op3}) and has a slower decreasing than the case in Fig. \ref{perc_op} obtained with the first definition of $T_{LE}$; as consequence, at each time step, the contribution to the energy density due to the electron--phonon part is almost constant for $LO$ and has a lower decreasing for optical isotropic phonons (Fig. \ref{W_ph3_pp}a) with respect to that obtained with the first definition of $T_{LE}$ (Fig. \ref{W_ph1_pp}a). Again, $\Delta W_{ep-op}$ is higher than $\Delta W_{ep-LO}$, sign that in the isotropic case electrons reach more energetic regions after collisions also with the second definition of $T_{LE}$. Nevertheless, in this case, as shown in Fig.  \ref{W_ph3_pp}(b), the phonon--phonon part is able to compensate the electron--phonon contribution and the mean values of net amounts $\Delta W$ with isotropy and anisotropy are comparable, as reported in Fig. \ref{W_ph3_pp}(c); the difference is small, and lower than that in Fig. \ref{W_ph1_pp}(c), explaining the reduction in the discrepancy between the isotropic and anisotropic local equilibrium temperatures $T_{LE}$ in  Fig. \ref{T_comp3}(a) when the second definition of local equilibrium temperature is taken into account.

\begin{figure}[h!]
	\centering
	\fbox {a)		\includegraphics[width=0.41\columnwidth]{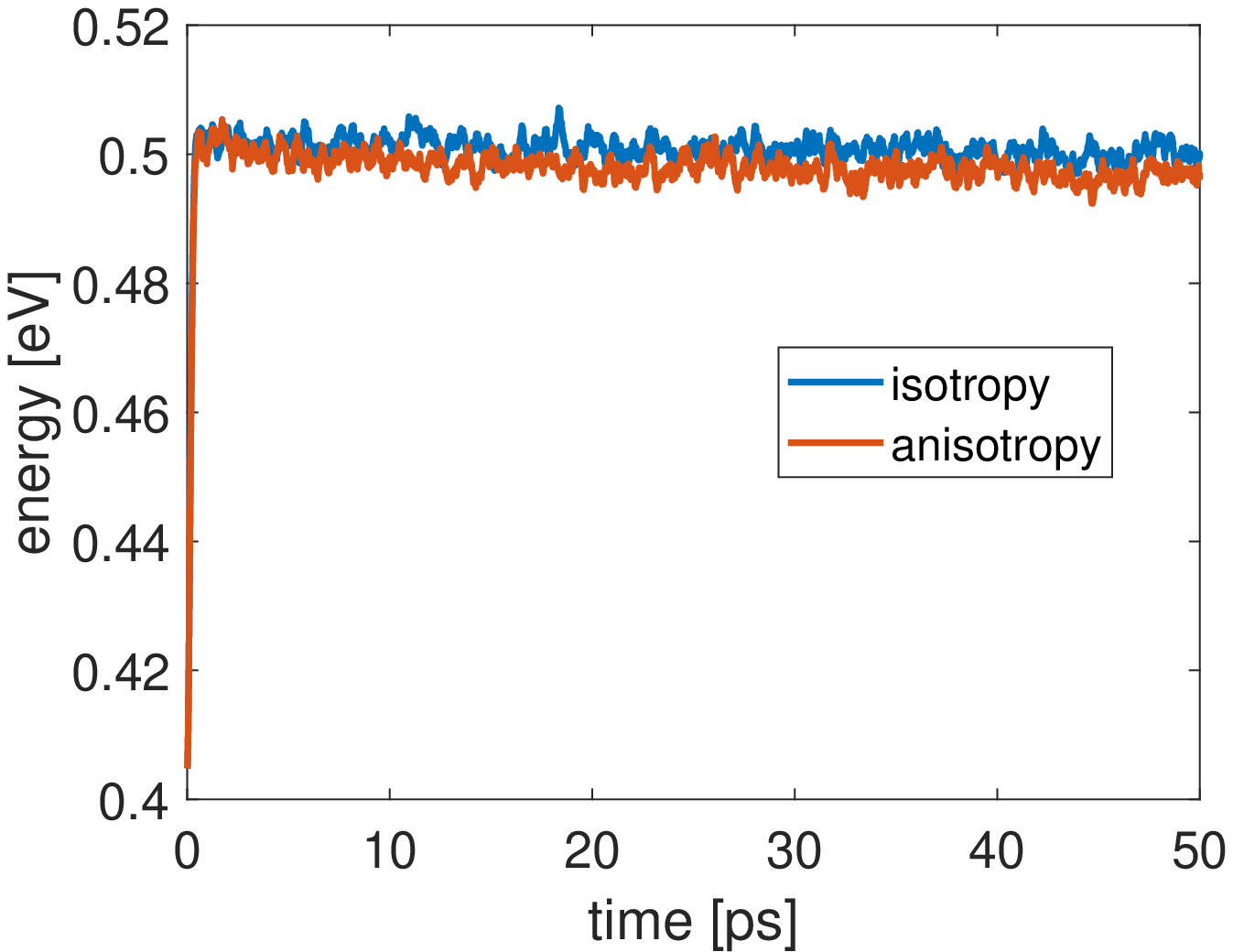}}
	\fbox {b)		\includegraphics[width=0.41\columnwidth]{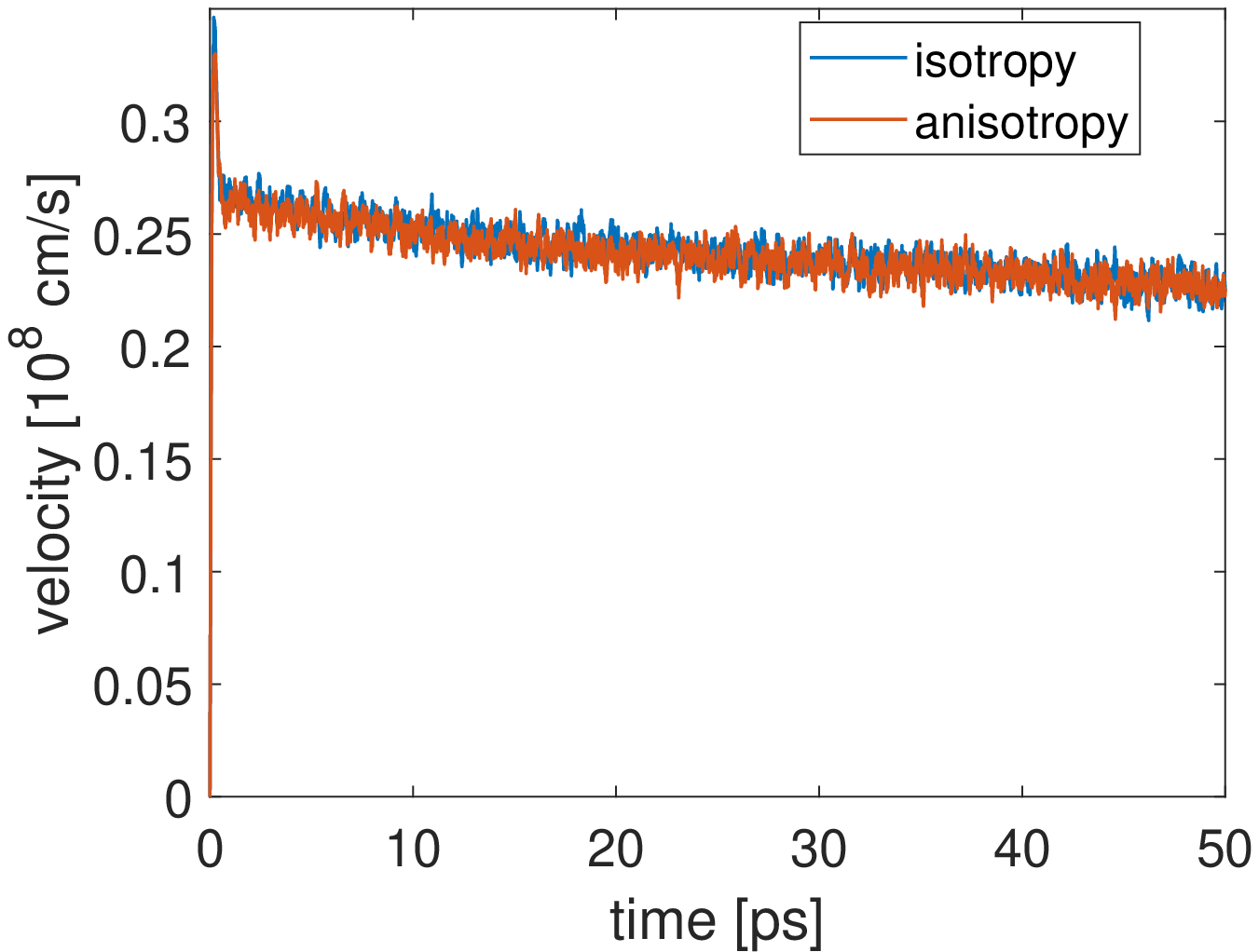}}\\
	\caption{Mean energy, (a), and velocity, (b), with and without anisotropy, with the second definition of $T_{LE}$. $\varepsilon_F=0.6~{\rm eV}$ and $E=20$ kV/cm.	\label{el_T2}}
\end{figure}

In Fig. \ref{el_T2}, average electron energy and velocity do not show any difference between the isotropic and anisotropic case when the second definition of $T_{LE}$ is used, at variance with results in Fig. \ref{el_T}. We highlight how mean energy in the isotropic approximation this time does not have any increasing from the steady state value. 

It is fundamental to note that the mean velocity at $50$ ps obtained in the anisotropic situation with the second definition of $T_{LE}$ is higher than that evaluated with the first definition of local equilibrium temperature (see Fig.s \ref{el_T}--\ref{el_T2} for comparison) and they differ of about 24\%. This result could be really useful in the debate regarding the correct definition of the graphene lattice local equilibrium temperature; it could be enough to measure the degradation of the electric current density when an external electric field is applied to experimentally identify which of the two definitions of $T_{LE}$ is correct, obviously assuming the anisotropic model the correct one; investigation of the graphene electrical behavior can be with this point of view an indirect experimental way to study its thermal properties, overcoming the difficulties related to direct thermal measurements.

\begin{figure}[h!]
	\centering
	\fbox {a)		\includegraphics[width=0.41\columnwidth]{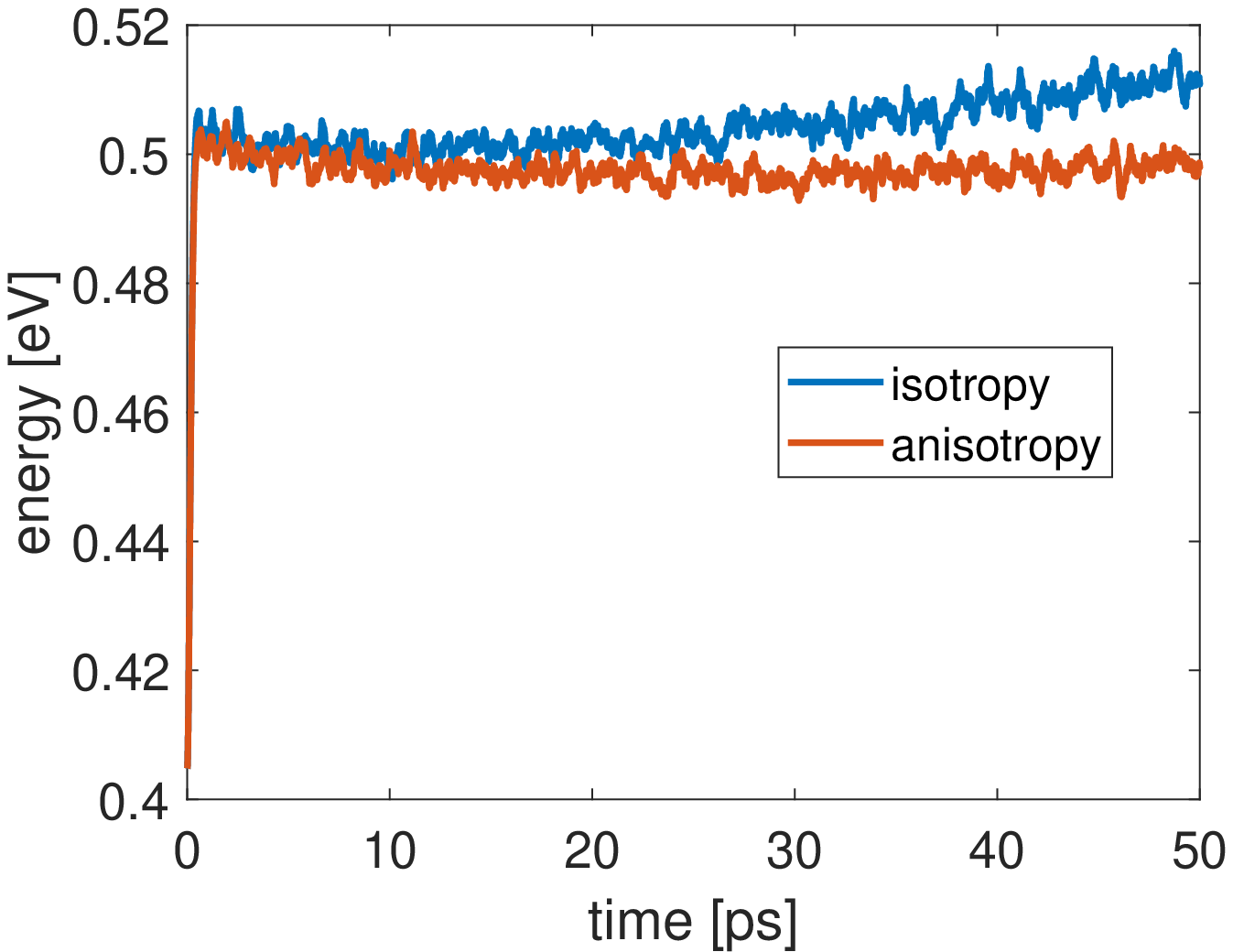}}
	\fbox {b)		\includegraphics[width=0.41\columnwidth]{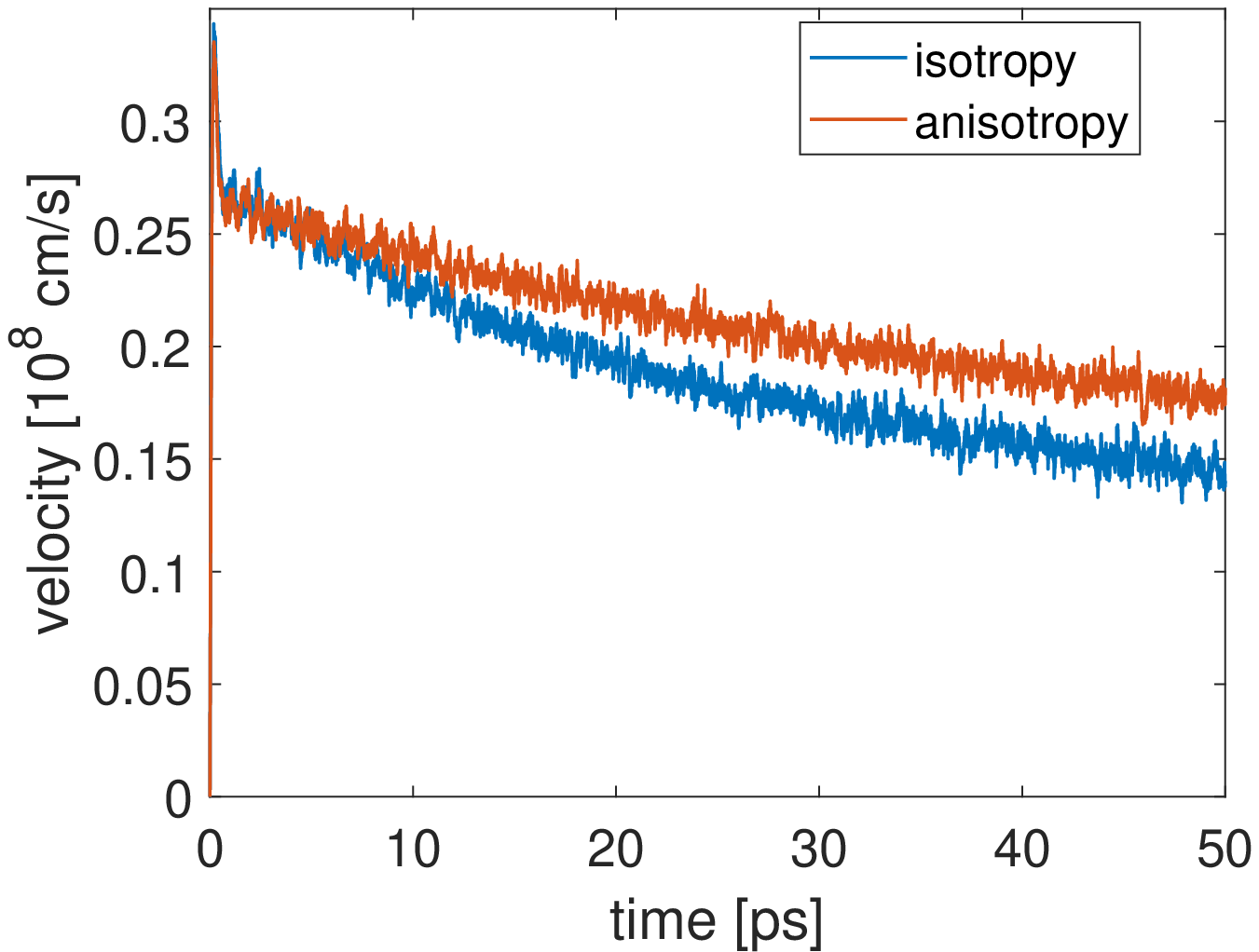}}\\
	\caption{Mean energy, (a), and velocity, (b), with and without anisotropy, with constant relaxation times, when the first definition of $T_{LE}$ is used. $\varepsilon_F=0.6~{\rm eV}$ and $E=20$ kV/cm.	\label{el_T1_tau}}
\end{figure}

\begin{figure}[h!]
	\centering
	\fbox {a)		\includegraphics[width=0.41\columnwidth]{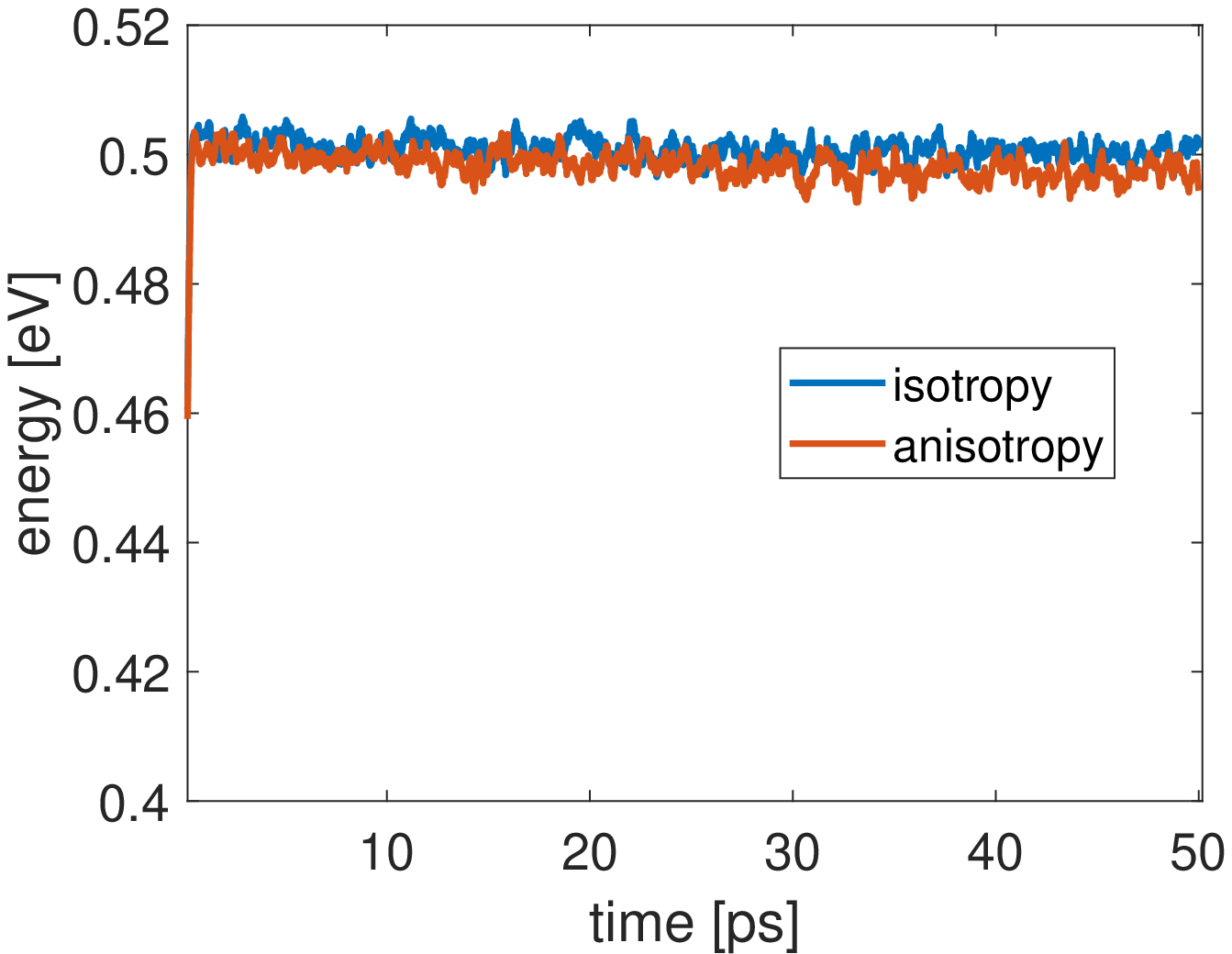}}
	\fbox {b)		\includegraphics[width=0.41\columnwidth]{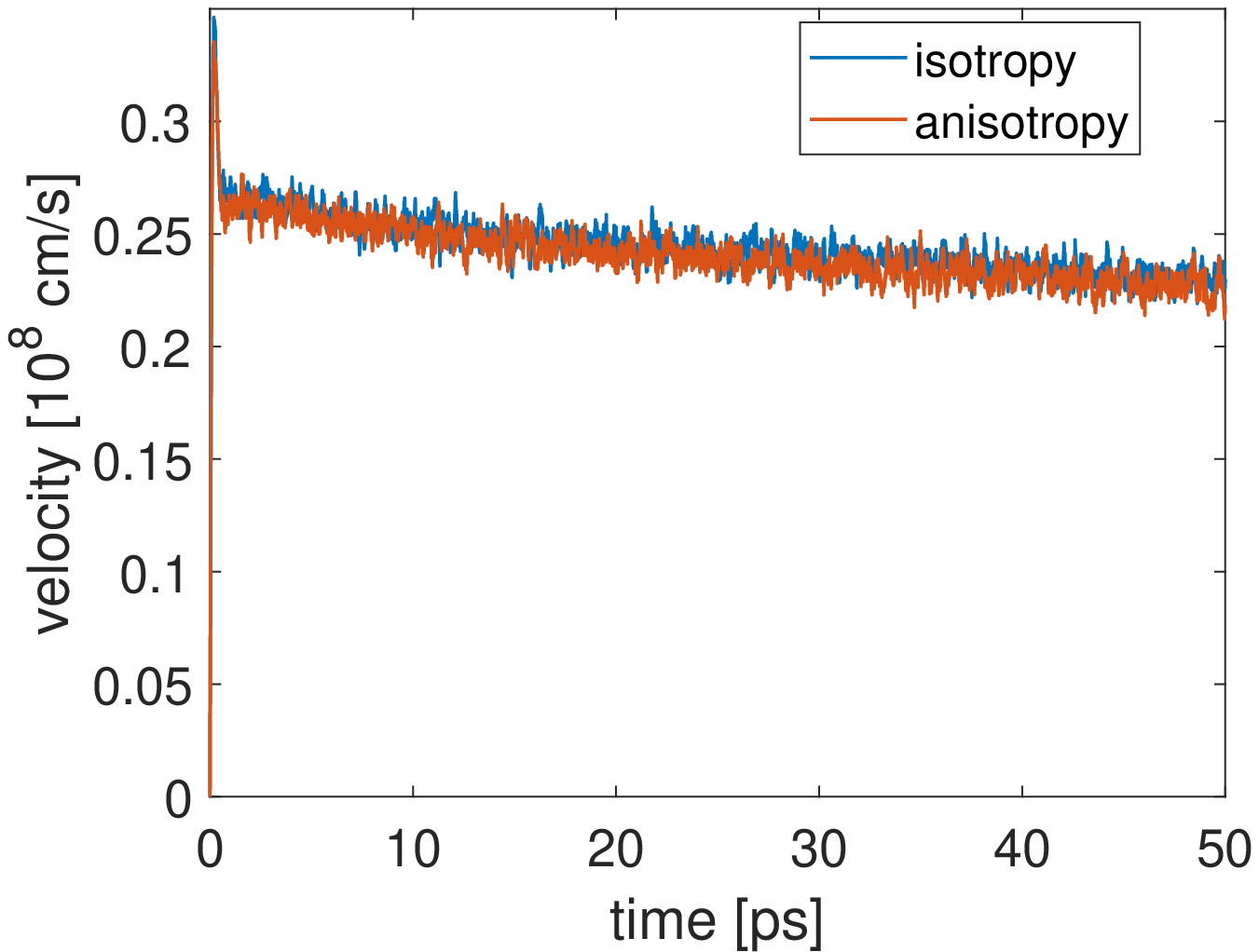}}\\
	\caption{Mean energy, (a), and velocity, (b), with and without anisotropy, with constant relaxation time, when the second definition of $T_{LE}$ is used. $\varepsilon_F=0.6~{\rm eV}$ and $E=20$ kV/cm.	\label{el_T2_tau}}
\end{figure}

A possible drawback could be given by the current lack of a definitive definition and numerical data of phonon relaxation times. Luckily, they have great importance on the evaluation of temperatures but it seems that they do not significantly alter electric characteristic curves. In Fig.s \ref{el_T1_tau}--\ref{el_T2_tau} we show mean electron energy and velocity obtained with the previous two definitions of local equilibrium temperature but by using constant relaxation times $\tau_{\mu}=1.5 \, \mbox{ps}$, except for $ZA$ phonons which maintain their already used relaxation time to avoid the introduction of numerical errors due to its relevant variation with respect to temperature. Numerical results about mean energy and velocity obtained with constant relaxation times are almost equal to the those evaluated with temperature dependent relaxation times (compare Fig.s \ref{el_T}--\ref{el_T2} and Fig.s \ref{el_T1_tau}--\ref{el_T2_tau}); above all, anisotropic electron mean velocity shows the same discrepancy between the values computed with the two definitions of $T_{LE}$ also with constant relaxation times, making the previous conclusion on the possibility to have an alternative experimental way to detect local equilibrium temperature in graphene more solid.

\section{Conclusions}
\label{Conclusions}

The first aim of this paper is to evaluate the influence of the planar phonon anisotropy in the study of charge and phonon transport in graphene. The numerical simulation procedure is hybrid, i.e.~stochastic and deterministic; electron transport is investigated with Direct Simulation Monte Carlo technique whose results enter as source terms in the phonon Boltzmann equations which are solved by means of numerical deterministic methods.

Although the isotropic approximation for planar phonons is often accepted for simulation purposes, an exhaustive analysis of its effect on numerical results for graphene is not present, at best of our knowledge, in the literature.

We discussed that isotropic approximation leads to some physically questionable results. Local equilibrium temperature and the phonon branches temperatures are highly overestimated; electron mean energy tends to increase from a steady state value, and with planar phonons' isotropy the charge mean velocity reaches considerably lower value than that in anisotropic case, when the definition of local equilibrium temperature based on the phonon--phonon collision operator properties is used; the difference between the isotropic and anisotropic electron mean velocity is negligible with the second definition of $T_{LE}$. Moreover, we distinguished the relative contribution of in-plane acoustic and optical phonons anisotropy; these last are the most effective. The out-of-plane $Z$ phonons confirmed they fundamental role in the analysis of thermal transport in graphene because, even if they do not directly interact with electrons, they are not negligible for the correct evaluation of temperatures.

Taking advantage of Monte Carlo procedures of being able to look into each physical step of the simulated phenomenon, {{we analyzed the contributions of the electron--phonon and phonon--phonon interactions to the phonon energy density.}} In the isotropic approximation, phonons have a higher net amount of transfers of quanta of energy to electrons than with anisotropy, but their temperatures and the local equilibrium temperature result higher; this behavior was explained by geometric considerations because angles between the charge wave vectors of initial and final state disappear in the scattering matrix elements for optical phonons, Eq.~\eqref{GLO_is}, when anisotropy is neglected, and electrons are allowed to reach more energetic regions with respect to the anisotropic case, in which geometric selection rules are present. Besides, with isotropy, charge carriers tend to acquire more often final negative velocities.

These results make planar phonon anisotropy not negligible even for simulation aims only. 

The present Monte Carlo procedure, able to correctly deal with the Pauli exclusion principle, and the coupling of the deterministic solution of the phonon Boltzmann equations with stochastic intermediate results, as explained in Section \ref{Model}, together with the discussed modeling of planar phonon anisotropy, constitute a coherent and self-consistent simulation tool for the study of charge and phonon transport in graphene in the semiclassical approximation.

This work would like to contribute also to the open debate about the definition of the crystal lattice local equilibrium temperature $T_{LE}$ in graphene. There are two main definitions in the literature described by Eqs. \eqref{prod_ph}--\eqref{second_T}.

We compared the simulation results obtained by using the two definitions of $T_{LE}$. We remark that the first definition given by Eq.~\eqref{prod_ph} contains explicitly the temperature dependent relaxation times, and in turn the anisotropy, which instead do not enter the Eq.~\eqref{second_T}. 

With the second definition of $T_{LE}$, the difference between the number of emission and absorption processes is comparable for isotropic and anisotropic cases; in the isotropic approximation, electrons tend to reach more energetic regions and geometric rules result again important but phonon--phonon collisions are more able to compensate the energy contribution arising from electron--phonons interactions, as discussed in the previous section. Nevertheless, temperatures are higher in the isotropic approximation with the second definition of $T_{LE}$ as well, although the discrepancy between the isotropic and anisotropic results is smaller than that observed with the first definition.

The contribution on the identification of the local equilibrium temperature arises from the analysis of the electron mean velocity. It decreases both when the first definition is used and with the second one, but with the first definition the final value at $50$ ps is lower for the anisotropic case of about 24\% with respect to that obtained with the second definition of $T_{LE}$. In this way, measurements of the degradation of the electric current density could be an experimental alternative method to detect the actual local equilibrium temperature in graphene, and to establish which of the two Eqs. \eqref{prod_ph}--\eqref{second_T} is experimentally confirmed. Such an approach could lead to overcome the current practical difficulties about direct thermal measurements on graphene.

The achievements of this paper may have positive implications in a wide range of applications, above all those related to the study and simulation of electronic devices, but also in physical and engineering fields which need similar computational tools. The correct evaluation of the crystal lattice local equilibrium temperature is fundamental for the theoretical studies about graphene and also for the applications, because it strongly influences the design of cooling mechanisms.

\section{Acknowledgments}
The author acknowledges the support of the {INdAM National Group of Mathematical Physics (GNFM, Italy)} and the financial support from project {FSE REACT-EU, Asse IV - Istruzione e ricerca per il recupero, Azione IV.6 - Contratti di ricerca su tematiche Green}, DM 1062 del 10-08-2021, ``Modelli fisico matematici per lo studio e la simulazione di sistemi e prodotti eco-sostenibili basati su nuovi materiali''.

\end{document}